\documentclass[a4paper,10pt,oneside,onecolumn,preprint,sort&compress]{elsarticle}
\usepackage{amsmath}
\usepackage{amssymb}

\title{Bias, variance, and confidence intervals for efficiency estimators in particle physics experiments}
\author[1]{Hans Dembinski\corref{cor1}}
\ead{Hans.Dembinski@tu-dortmund.de}
\author[2]{Michael Schmelling}
\ead{Michael.Schmelling@mpi-hd.mpg.de}
\cortext[cor1]{Corresponding author}

\address[1]{TU Dortmund, Otto-Hahn Str. 4a, 44221, Dortmund, Germany}
\address[2]{Max-Planck-Institut für Kernphysik, Saupfercheckweg 1, 69117 Heidelberg, Germany}

\usepackage{xspace}
\usepackage{ifthen}

\graphicspath{{fig/}}

\newcommand{\fig}[3][width=0.9\textwidth]{\begin{figure}\centering\includegraphics[#1]{#2}\caption{#3}\label{fig:#2}\end{figure}}

\newcommand{\avg}[1]{\langle #1 \rangle}

\newcommand{\fg}[1]{Fig.~\ref{fig:#1}\xspace}

\newcommand{\eq}[1]{Eq.~\ref{eq:#1}\xspace}

\newcommand{\dd}{\text{d}}

\newcommand{\beq}[1][]{%
  \begin{equation}%
  \ifthenelse{\equal{#1}{}}{
  }{\label{eq:#1}}
}
\newcommand{\eeq}{\end{equation}}
\newcommand{\eql}[1]{\label{eq:#1}}

\DeclareMathOperator{\var}{var}
\DeclareMathOperator{\ExOp}{E}

\newcommand{\Ex}[1]{\ExOp\!\left[#1\right]}

\newcommand{\probp}{\text{Pois}}
\newcommand{\probb}{\text{B}}

\newcommand{\sigb}[1]{\sigma_{#1,b}}

\begin{document}

\begin{abstract}
    We compute bias, variance, and approximate confidence intervals for the efficiency of a random selection process under various special conditions that occur in practical data analysis. We consider the following cases: a) the number of trials is not constant but drawn from a Poisson distribution, b) the samples are weighted, c) the numbers of successes and failures have a variance which exceeds that of a Poisson process, which is the case, for example, when these numbers are obtained from a fit to mixture of signal and background events. Generalized Wilson intervals based on these variances are computed, and their coverage probability is studied. The efficiency estimators are unbiased in all considered cases, except when the samples are weighted. The standard Wilson interval is also suitable for case a). For most of the other cases, generalized Wilson intervals can be computed with closed-form expressions.
\end{abstract}

\maketitle

\section{Introduction}

In (astro)particle physics experiments, a cross-section or a flux is computed from the number of times a particular type of event occurs within a time-frame with an associated integrated luminosity or exposure. This is essentially a rate measurement, where the integrated luminosity or exposure plays of the role of time. The events of interest are usually observed imperfectly with an efficiency smaller than one. The efficiency $p$ of a detector or a selection is
\begin{equation}
    p = \frac{n_1}{n_1 + n_2} = \frac{n_1} n,
\end{equation}
where $n_1$ is the number of events which are selected, $n_2$ is the number of events that are rejected, and $n$ is the total number of trials. The value of $p$ is usually estimated either from Monte-Carlo simulation or calibration samples, using the estimator $\hat p$, obtained by replacing the true values $n_k$ with their sample values $\hat n_k$,
\begin{equation}
    \hat p = \frac{\hat n_1}{\hat n_1 + \hat n_2}.
\end{equation}
If the true efficiency $p$ is constant and $\hat n_1 + \hat n_2 = n$ is fixed, then $\hat n_1$ is binomially distributed. Since $\hat p$ is a linear function of $\hat n_1$, $\hat p$ is an unbiased estimate of $p$,
\begin{equation}
    \Ex {\hat p} = \frac{\Ex{\hat n_1}}{n} = \frac{n_1}{n} = p,
\end{equation}
and the variance of $\hat p$ can be calculated exactly to
\begin{align}
    \var(\hat p) & = \Ex{(\hat p - p)^2} \nonumber                                                                              \\
                 & = \sum_{\hat n_1 = 0}^\infty \left( \frac{\hat n_1}{n} - p \right)^2 \probb(\hat n_1; n, p)
    = \frac 1 {n^2}
    \sum_{\hat n_1 = 0}^\infty (\hat n_1 - p\, n)^2 \probb(\hat n_1; n, p) \nonumber                                            \\
                 & = \frac{\Ex{(\hat n_1 - n_1)^2}} {n^2} = \frac{p (1 - p)} n = \frac{n_1 n_2}{(n_1 + n_2)^3}, \eql{var_binom}
\end{align}
where we used the known second central moment of the binomial distribution $\probb(n, p)$.

This binomial model is the usual starting point on which confidence intervals for the efficiency estimator are derived. Confidence intervals are constructed so that they contain the true value with a fixed probability in repeated identical experiments, the coverage probability.

We emphasize that coverage probability is not only a theoretical concept, but also a useful concept in practice. We illustrate this with the common case where the efficiency $p$ depends on some variable $x$, and efficiency estimates are computed in bins of $x$. In general, the true values of $p_i$ and $n_i$ in each bin $i$ differ, and thus each bin can be regarded as an independent experiment. The actual result in each bin is one sample from the respective probability distributions. By constructing intervals for each bin that have the same known coverage probability, we guarantee that the whole set of intervals over all bins also has that coverage probability to cover the respective true values $p_i$. This allows one to judge whether a model for the $p(x)$ agrees with the measurements. If it does, the fraction of the intervals that intersect with the curve $p(x)$ approaches the coverage probability (for standard intervals this is the well-known \emph{rule of 2/3}).

A universally accepted formula for the confidence interval of an efficiency is not at hand and many formulas are discussed in the literature. A comprehensive review is given by Cousins, Hymes, and Tucker \cite{Cousins:2009kz}. Because of the discreteness of the counts, it is not possible to construct intervals which contain the true value with a fixed probability for all true values. Clopper-Pearson intervals \cite{10.1093/biomet/26.4.404} always contain the true value with at least the given probability, but overcover on average (for any kind of averaging). When efficiencies are computed as intermediate results in a data analysis, using Clopper-Pearson intervals is inconsistent with the general approach of choosing intervals that neither over- nor undercover on average, since this would lead to violations of the rule of 2/3. In the example from the previous paragraph, the fraction of Clopper-Pearson intervals that overlap with the model curve $\hat p(x)$ would be always larger than 68\,\%. Other intervals have been proposed, which have an average coverage probability closer to the nominal value for randomly chosen values of $p$ and $n$, at the cost of undercovering for some values. We will focus on the Wilson interval \cite{doi:10.1080/01621459.1927.10502953}, because it has attractive properties and is easy to generalize. The Wilson interval was also recommended in Ref.\,\cite{Cousins:2009kz} when being conservative is not the primary goal.

In this work, we investigate several cases that commonly appear in practice and which go beyond the binomial model. Firstly, we note that a fixed number of trials $n$ is an improper model for particle physics experiments. The actual number of trials $\hat n$ in most experiments is Poisson distributed, because experiments are run for a fixed time rather than until a fixed number of trials has been accumulated. This modifies the variance of $\hat p$ compared to what is expected from the binomial model. Since the Wilson interval is based on this variance, we investigate the impact of the modified variance on the Wilson interval, but find that the standard Wilson interval is still suitable for this case. We further consider weighted samples, in which the number of successes and failures are sums of weights, and situations in which the counts $\hat n_k$ are not directly known, but replaced with estimates that have increased variance compared to Poisson fluctuations. We compute the bias and variance of $\hat p$ under each condition and generalized Wilson intervals for all cases, that are based on simple closed formulas.

\section{Poisson-distributed trials}

We are considering the case where $p$ is a fixed number and the actual number of trials $\hat n$ is Poisson-distributed around an expectation $n$. There are two equivalent ways to calculate the probability for observing a particular value of $\hat p$ in this case. The first is based on the straight-forward product of the probabilities of the binomial distribution $\probb(n, p)$ and the Poisson distribution $\probp(n)$, while the second uses the probabilities from two Poisson distributions. The probability to observe the sample pair $\hat n_1, \hat n$ is
\begin{align}
    P(\hat n_1, \hat n; p, n)
     & = \probb(\hat n_1; \hat n, p) \probp(\hat n; n) \nonumber                                                                             \\
     & = \frac{\hat n!}{\hat n_1! (\hat n - \hat n_1)!} p^{\hat n_1} (1 - p)^{\hat n - \hat n_1} \frac{e^{-n} n^{\hat n}}{\hat n!} \nonumber \\
     & = \frac{1}{\hat n_1! \hat n_2!} p^{\hat n_1} (1 - p)^{\hat n_2} e^{-n_1 - n_2}  n^{\hat n_1 + \hat n_2} \nonumber                     \\
     & = \frac{e^{-n_1} n_1^{\hat n_1}}{\hat n_1!}  \frac{e^{-n_2} n_2^{\hat n_2}}{\hat n_2!} \nonumber                                      \\
     & = \probp(\hat n_1; n_1) \probp(\hat n_2; n_2),
\end{align}
with $n_1 = p n$ and $n_2 = n - n_1 = (1- p) n$. Both forms are equivalent, so one can use the most convenient form for each calculation.

We calculate the bias and variance of the efficiency estimator $\hat p$ again under these new conditions. The estimator is undefined for $\hat n = 0$, so one has to decide how this case should be handled. We could, for example, assign the value $1/2$ to $\hat p$, but that would not correspond to actual practice. Instead, we skip these outcomes with $\hat n = 0$ and compute the expectation of the remaining cases. This leads to some surprising properties, as we will see.

Calculating the bias is important since the estimator $\hat p$ is now a non-linear function of the arguments $\hat n_1$ and $\hat n$. Non-linear functions in general lead to biased estimators, but $\hat p$ is still unbiased. To show this, we compute the expectation of $\hat p$,
\begin{align}
    \Ex{\hat p} & = \frac{1}{1 - \probp(0; n)} \sum_{\hat n=1}^\infty \sum_{\hat n_1=0}^\infty \frac{\hat n_1}{\hat n} \probb(\hat n_1; \hat n, p) \, \probp(\hat n; n) \nonumber \\
                & = \frac{1}{1 - \probp(0; n)} \sum_{\hat n=1}^\infty \frac{\hat n p}{\hat n} \probp(\hat n; n) = p.
\end{align}
Next, we compute the variance of $\hat p$ and get
\begin{align}
    \var(\hat p) & = \Ex{(\hat p - p)^2} \nonumber                                                                                                                                                     \\
                 & = \frac{1}{1 - \probp(0; n)} \sum_{\hat n = 1}^\infty \sum_{\hat n_1 = 0}^\infty \left(\frac{\hat n_1}{\hat n} - p\right)^2 \probb(\hat n_1; \hat n, p) \probp(\hat n; n) \nonumber \\
                 & = \frac{1}{1 - \probp(0; n)} \sum_{\hat n = 1}^\infty \frac{p (1 - p)}{\hat n} \probp(\hat n; n) \nonumber                                                                          \\
                 & = \frac{p(1 - p)}{n} \frac{1}{1 - \probp(0; n)} \sum_{\hat n = 1}^\infty \frac{n}{\hat n} \probp(\hat n; n) \nonumber                                                               \\
                 & = \frac{p (1 - p)}{n} f(n) . \eql{var_poisson}
\end{align}
We used \eq{var_binom} to replace the sum over $\hat n_1$. In the last step, we expressed the variance of $\hat p$ as the product of \eq{var_binom} with a correction $f(n)$. The calculation of $f(n)$ can be carried out numerically. Terms in which $\hat n$ deviates a lot from $n$ quickly approach zero and thus only a few terms of the sum around the expectation value $n$ need to be evaluated. Alternatively, one can compute the variance based on Monte-Carlo simulation, by computing the average $\avg{(\hat p - p)^2}$ over many randomly drawn samples. Due to law of large numbers, this approaches $\Ex{(\hat p - p)^2}$.

\fig{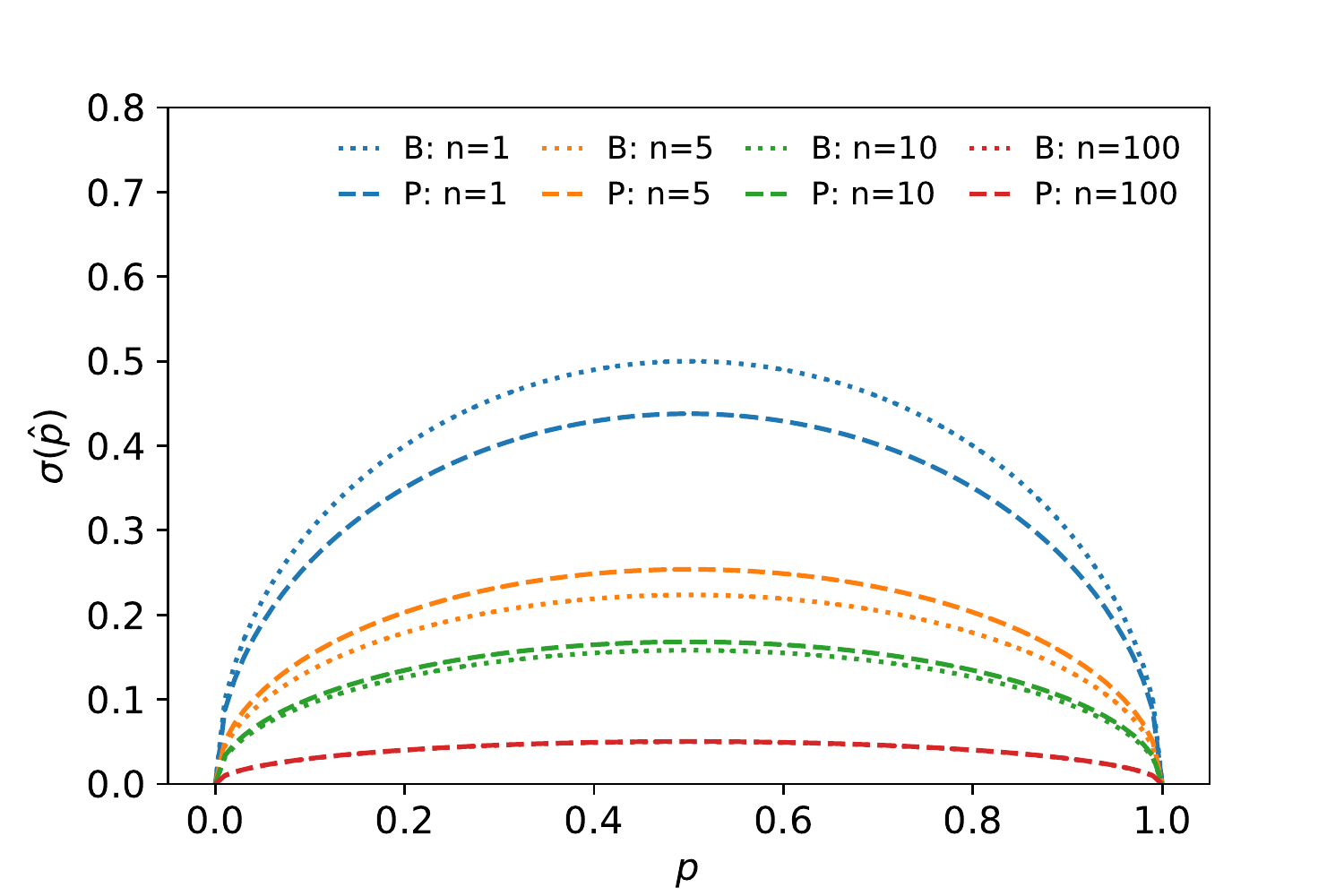}{Standard deviation (square-root of the variance) of the efficiency estimator $\hat p$ as a function of the true efficiency $p$ for different values of $n$. Shown are results for the binomial distribution when $\hat n = n$ is fixed (dotted lines) and when $\hat n$ is Poisson distributed with expectation $n$ (dashed lines).}

\fig{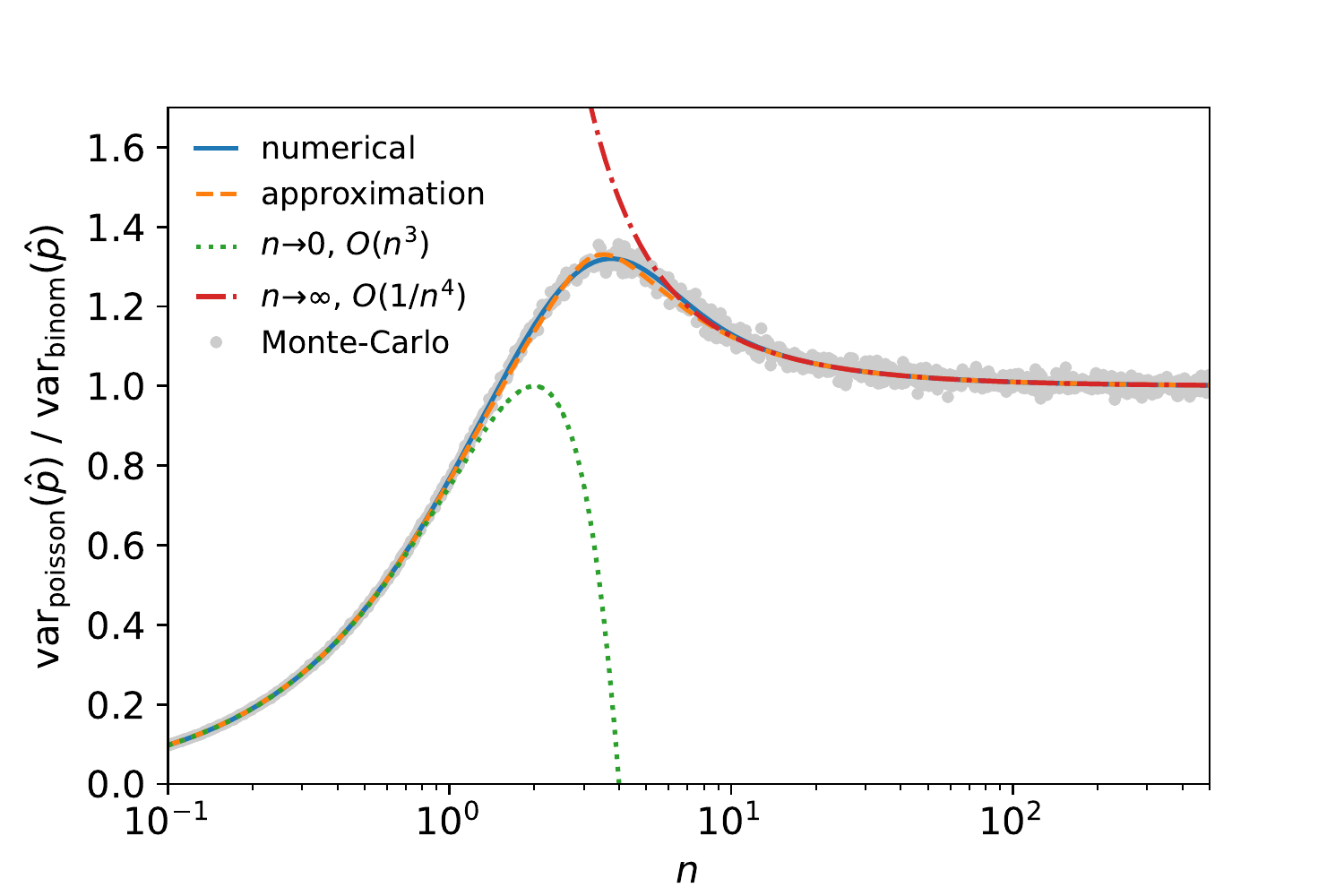}{Variance of the efficiency estimator $\hat p$ when $n$ is Poisson distributed as a function of the expectation value of $n$ divided by $\var_\text{binom}(\hat p)$. Shown are the numerical computation (solid line), the Taylor series for $n \to \infty$ up to $O(1/n^4)$ (dotted line), the Taylor series for $n \to 0$ up to $O(n^2)$ (dash-dotted line), and the approximate computation described in the text (dashed line). Also shown are results from a Monte-Carlo simulation (dots).}

Results of solving \eq{var_poisson} numerically are shown in \fg{standard_deviation_comparison} and compared to results from \eq{var_binom}. The results agree in the limit $n \to \infty$, but differ for $n \lesssim 10$. The correction $f(n)$ is shown in \fg{variance_ratio} for a numerical and a Monte-Carlo calculation of \eq{var_poisson}. Its starts off smaller than one, due to the exclusion of samples with $\hat n = 0$, and then rises above one with a maximum in the range $n = 3 \dots 4$ and then approaches one as $n$ increases further.

Since \eq{var_poisson} is still comparably expensive to compute, we investigate some approximations that are quick to compute. We start by expanding the estimator $\hat p = \hat n_1 / (\hat n_1 + \hat n_2)$ at $\hat n_k = n_k$ in a Taylor series,
\begin{equation}
    \hat p = \sum_{i=0}^\infty \sum_{j=0}^\infty \frac{(\hat n_1 - n_1)^i}{i!} \frac{(\hat n_2 - n_2)^j}{j!} \frac{\partial^{i+j} \hat p}{\partial \hat n_1^i \partial \hat n_2^j}(n_1, n_2),
\end{equation}
treating the integer values $\hat n_k$ like real numbers. This expansion approximates the asymptotic limit $n_k \to \infty$. We can rewrite this as
\begin{multline}
    \hat p - p =
    \sum_{j=1}^\infty \frac{(\hat n_2 - n_2)^j}{j!} \frac{\partial^j \hat p}{\partial \hat n_2^j} (n_1, n_2) +
    \sum_{i=1}^\infty \frac{(\hat n_1 - n_1)^i}{i!} \frac{\partial^i \hat p}{\partial \hat n_1^i} (n_1, n_2) \; + \\
    \sum_{i=1}^\infty \sum_{j=1}^\infty \frac{(\hat n_1 - n_1)^i}{i!} \frac{(\hat n_2 - n_2)^j}{j!} \frac{\partial^{i+j} \hat p}{\partial \hat n_1^i \partial \hat n_2^j}(n_1, n_2). \eql{taylor}
\end{multline}
The variance of $\hat p$ can then be computed by squaring both sides and computing the expectation on both sides. The expectation is a linear operator and therefore distributive to each term of the sum on the right-hand side. These terms consist of constants multiplied with
\begin{equation}
    \Ex{(\hat n_1 - n_1)^i (\hat n_2 - n_2)^j} = \Ex{(\hat n_1 - n_1)^i} \Ex{(\hat n_2 - n_2)^j}.
\end{equation}
The expectations of $\hat n_1$ and $\hat n_2$ factor, because they are independently distributed. The expectation $\Ex{(\hat n_k - n_k)^j}$ is the $j$-th central moment of the Poisson distribution, which can be computed to any order from the moment generating function. Putting all this together, the correction to first order in $n^{-1}$ is
\begin{equation}
    f_{n \to \infty}(n) = \frac{n + 1}{n} + O(n^{-2}).
\end{equation}
Although the simplicity of this formula is attractive, the accuracy is not very good. A correction to third order performs better,
\begin{equation}
    f_{n\to \infty}(n) = \frac{2n + n^2 + n^3 + 6}{n^3} + O(n^{-4}),
    \eql{fn_o4}
\end{equation}
and produces accurate results for $n \gtrsim 5$. It fails to describe the true variance as $n$ approaches zero, however. Another expansion of \eq{var_poisson} around $n_1 = n_2 = 0$ to second order yields
\begin{equation}
    f_{n \to 0}(n) = n + n^2 / 4 + O(n^3).
\end{equation}
We combined the two limits empirically with a transition function to obtain an approximation with an accuracy better than 1.7\,\% over the whole range,
\begin{gather}
    f_\text{approx}(n) = \big(1 - z(n)\big) \, \left(
    n - \frac{n^2}4
    \right)
    + z(n) \left(
    \frac{2n + n^2 + n^3 + 6}{n^3}
    \right) \eql{fn_approx} \\
    \text{with } z(n) = 1/(1 + e^{-(\ln_{q=0.82}(n) - \ln_{q=0.82}(2.92)) / 0.18} ), \nonumber
\end{gather}
in which we use the $q$-logarithm $\ln_q(x) = (x^{1-q} - 1) / (1 - q)$ for $q \neq 1$. The approximation and the two limiting functions are shown in \fg{variance_ratio}. We found a speed-up factor of about 500 in a comparison of a naive implementation that solved \eq{var_poisson} numerically with this approximation.

\section{Weighted samples}

Efficiencies are also often estimated from weighted samples. Weighting is used in particle physics to refine a sample obtained from Monte-Carlo simulation of an experiment to better reflect the frequencies of events with certain outcomes in the real experiment. The weights are frequency ratios in this case, non-negative, and not random numbers.

We investigate the bias and variance of the estimator $\hat{p}$,
\begin{equation}
    \hat p = \frac{\sum_i w_i}{\sum_\ell^{\hat n} w_\ell},
    \eql{phat_w}
\end{equation}
where the sum in the numerator goes over the weights of successes and the sum in the denominator goes over all weights. We further assume that the total number of trials $\hat n$ follows a Poisson-distribution, as before.

The following results are correct for any weight distribution if the weights are sampled independently of the successes and failures. Negative weights are allowed as long as the expectation $\Ex{w}$ is still positive, but \eq{phat_w} is less stable in this case and can produce values outside the interval $[0, 1]$.

We first discuss the simpler case when the weights are independent of the successes and failures and then discuss the case when both the weights and the success probability depend on a property of the event.

\subsection{Success probability independent of weight distribution}

The estimator $\hat{p}$ is unbiased with respect to $p$. To show this, we write it in the equivalent form
\begin{equation}
    \hat p = \frac{\sum_\ell^{\hat n} X_\ell w_\ell}{\sum_\ell^{\hat n} w_\ell},
    \eql{phat_w_alt}
\end{equation}
where the $X_{\ell} \in \{0, 1\}$ are drawn from Bernoulli distributions $\probb(1, p(x_\ell))$. We proceed to evaluate $\Ex{\hat p}$ sequentially. By taking the expectation over the Bernoulli distributions first, we see that the estimator is unbiased,
\begin{equation}
    \Ex{\hat p} = \Ex{\frac{\sum_\ell^{\hat n} p \, w_\ell}{\sum_\ell^{\hat n} w_\ell}} = p \Ex{\frac{\sum_\ell^{\hat n} \, w_\ell}{\sum_\ell^{\hat n} w_\ell}} = p.
\end{equation}

Next, we compute the variance of $\hat p$. We compute the expectation again sequentially and take the equation over the Bernoulli distribution first,
\begin{align}
    \Ex{(\hat p - p)^2}
     & = \Ex{\left(\frac{\sum_\ell^{\hat n} X_\ell w_\ell}{\sum_\ell^{\hat n} w_\ell} - p\right)^2}
    = \Ex{\left(\frac{\sum_\ell^{\hat n} (X_\ell - p) w_\ell}{\sum_\ell^{\hat n} w_\ell}\right)^2} \nonumber \\
     & = \Ex{\frac{\sum_\ell^{\hat n} (X_\ell - p)^2 w^2_\ell}{(\sum_\ell^{\hat n} w_\ell)^2}}
    = p(1 - p) \Ex{\frac{\sum_\ell^{\hat n} w_\ell^2}{(\sum_\ell^{\hat n} w_\ell)^2}}.
\end{align}
In the third step, we used that the expectations of mixed terms in the multinominal product disappear, since the $X_\ell$ are independently drawn, and we have
\begin{equation}
    \Ex{(X_i - p) (X_j - p)} = \Ex{X_i - p} \Ex{X_j - p} = 0 \text{ for } i \neq j,
\end{equation}
while we get for the quadratic terms
\begin{equation}
    \Ex{(X_i - p)^2} = p (1 - p).
\end{equation}
To calculate further, we consider the limit $\hat n \to \infty$, which allows us to use the law of large numbers to replace the sums in the fraction with expectations,
\begin{equation}
    \Ex{\frac{\sum_\ell^{\hat n} w_\ell^2}{(\sum_\ell^{\hat n} w_\ell)^2}}
    \xrightarrow{\hat n \to \infty} \frac{\hat n \Ex{w^2}}{(\hat n \Ex w)^2} = \frac{1}{\hat n} \frac{\Ex{w^2}}{\Ex{w}^2}.
    \eql{var_w_approx_inner}
\end{equation}
Finally, we get the asymptotic formula valid in the limit $n \to \infty$,
\begin{align}
    \var(\hat p) & = \Ex{(\hat p - p)^2} \nonumber                                                                                                                          \\
                 & \approx \frac{1}{1 - \probp(0; n)} \sum_{\hat n = 1}^\infty p(1-p) \frac{1}{\hat n} \frac{\Ex{w^2}}{\Ex{w}^2} \, \probp(\hat n; n) \nonumber             \\
                 & \approx \frac{p(1 - p)}{n} \frac{\Ex{w^2}}{\Ex{w}^2} \frac{1}{1 - \probp(0; n)} \sum_{\hat n = 1}^\infty \frac{n}{\hat n} \, \probp(\hat n; n) \nonumber \\
                 & = \frac{p(1 - p)}{n \frac{\Ex{w}^2}{\Ex{w^2}}} f(n) = \frac{p(1 - p)}{n_\text{eff}} f(n).
    \eql{var_w_approx}
\end{align}
In the last step, we introduced the effective count $n_\text{eff} = n \Ex{w}^2/ \Ex{w^2}$, and note that \eq{var_w_approx} resembles \eq{var_poisson} with $n$ replaced by $n_\text{eff}$. The formula is correct for large $n$, since contributions from small $\hat n$ become negligible where \eq{var_w_approx_inner} is not valid. To compute a sample estimate of the variance, one replaces $n_\text{eff}$ with its consistent estimate
\begin{equation}
    \hat n_\text{eff} = \frac{(\sum_i w_i)^2}{\sum_i w_i^2}.
    \eql{hat_neff}
\end{equation}

\begin{figure}
    \includegraphics[width=0.5\textwidth]{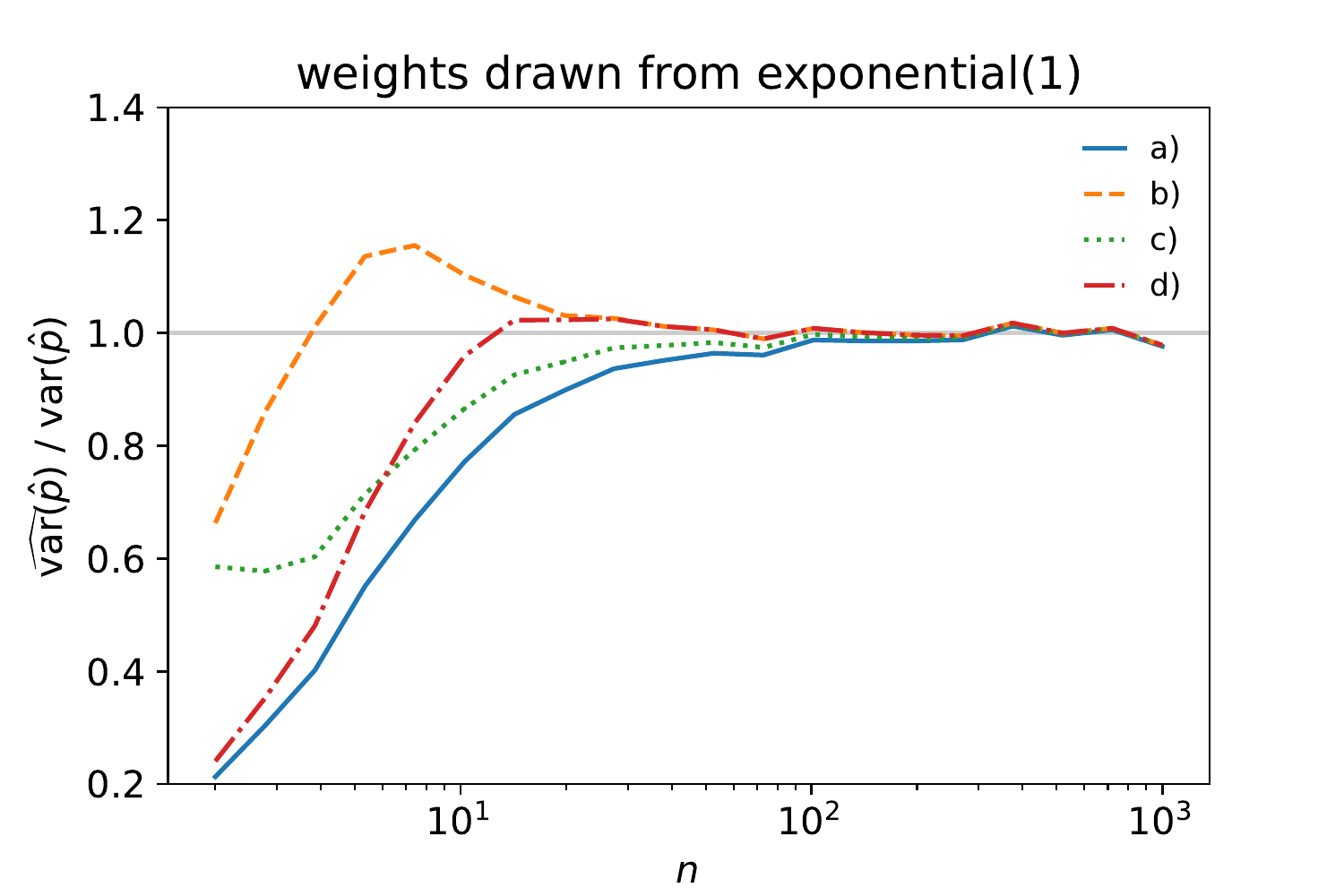}
    \includegraphics[width=0.5\textwidth]{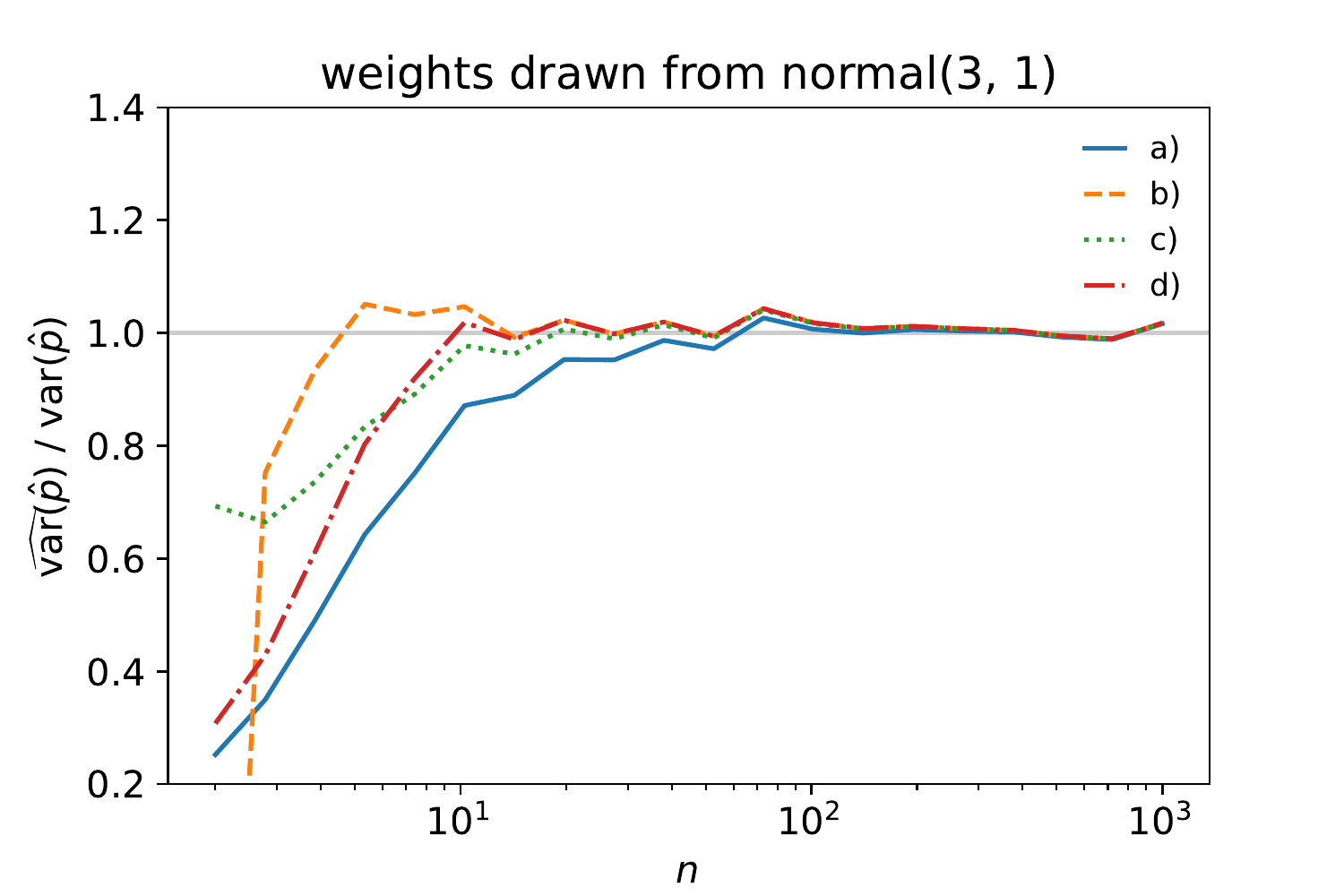}
    \includegraphics[width=0.5\textwidth]{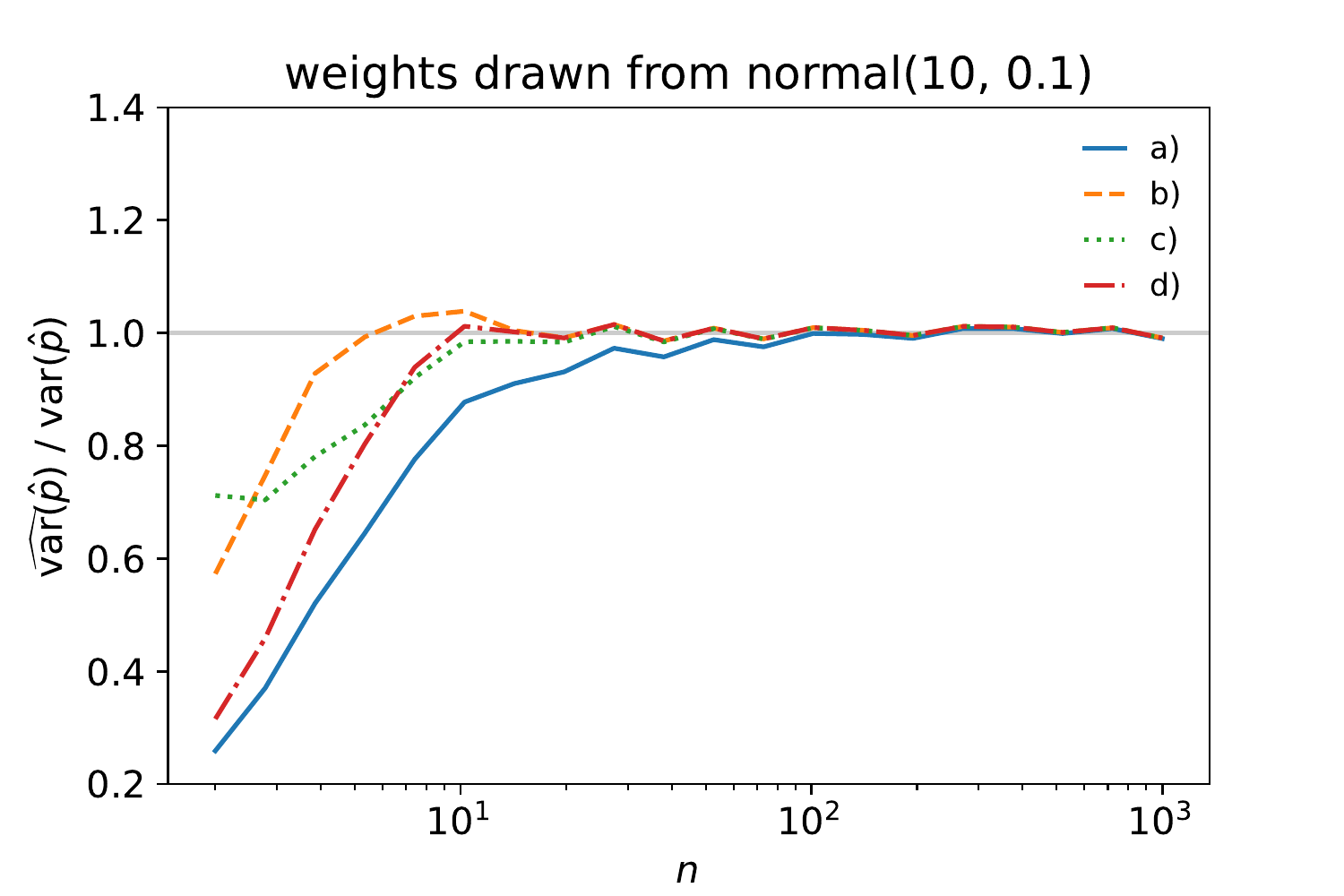}
    \includegraphics[width=0.5\textwidth]{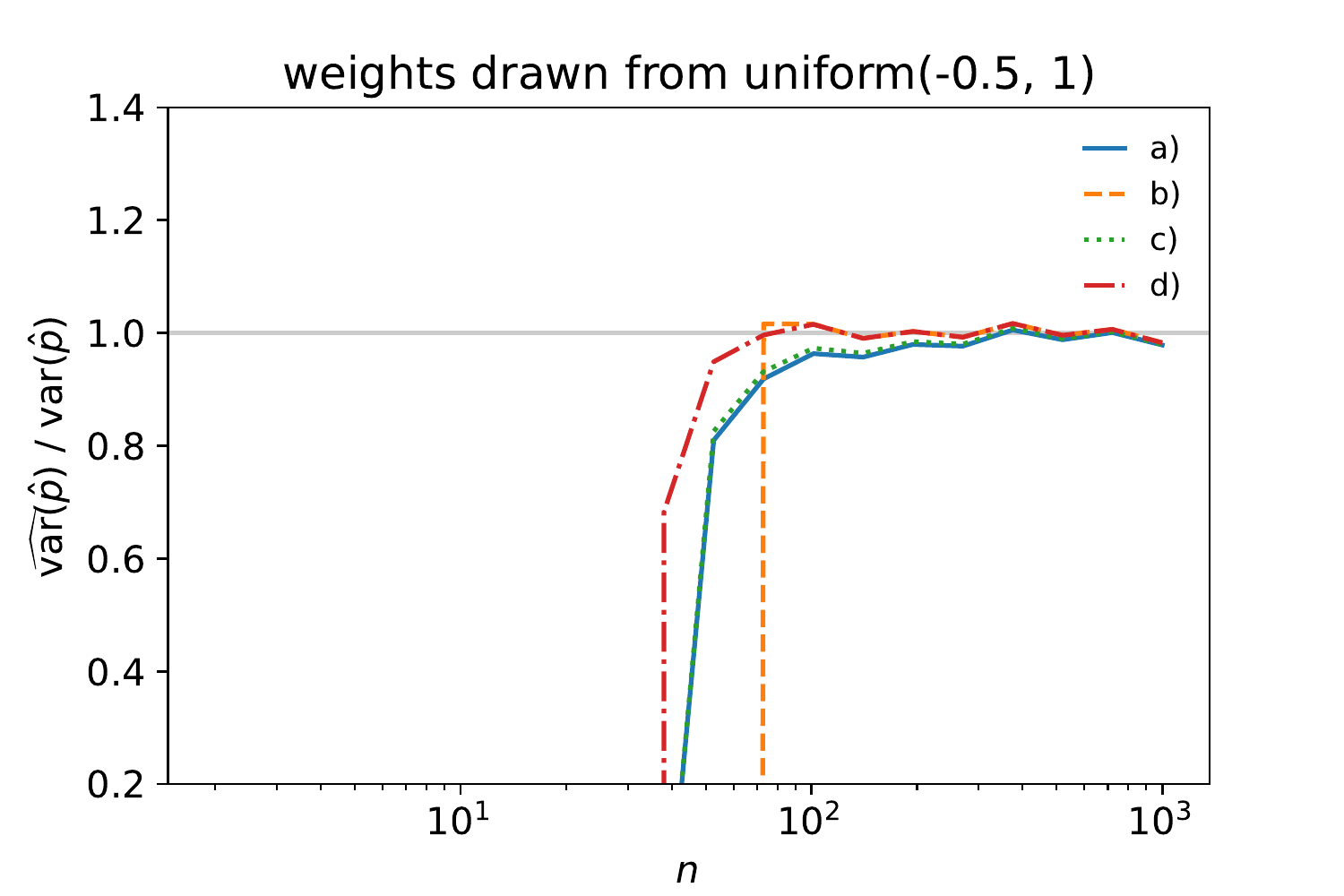}
    \caption{Average variance estimate of $\hat p$ divided by the variance from Monte-Carlo simulation for various weight distributions (exponential distribution with expectation 5, normal distributions with $(\mu = 3, \sigma = 1)$ and $(\mu = 10, \sigma = 0.1)$), and uniform distribution between $-0.5$ and $1$. The estimates are computed with \eq{var_w_approx} using a) $f(n) = 1$; b) $f(n)$ approximated by \eq{fn_o4}; c) $f(n_\text{eff})$ approximated by \eq{fn_o4}; d) $f(n_\text{eff})$ approximated by \eq{fn_approx}.}
    \label{fig:variance_weighted_sum}
\end{figure}

Although it is formally not correct to also replace $n$ with $n_\text{eff}$ in $f(n)$, this works very well in practice. We run toy simulations, shown in \fg{variance_weighted_sum}, in which the variance of $\hat p$ is computed for different weight distributions and compared with estimates based on \eq{var_w_approx}. For large $n$, all variants converge. The best approximation to the actual variance is given by \eq{var_w_approx} with $f(n_\text{eff})$ approximated by \eq{fn_o4}.

\subsection{Success probability and weight depend on event property}

The efficiency and weight often both depend on a property $x$ of the event. The goal is to compute the effective efficiency $\bar p$ obtained by integrating over $x$ with distribution $f(x)$,
\begin{equation}
    \bar p = \frac{\int w(x) \, p(x) \, f(x) \, \dd x }{\int w(x) \, f(x) \, \dd x } .
    \eql{pbar}
\end{equation}
In an example from high-energy physics, $p(x)$ represents the tracking efficiency, $x$ the track multiplicity in a bunch crossing, and $w(x)$ the ratio of the $x$-distributions of the real and simulated experiment. The weights are used to correct the $x$-distribution in the simulation. We take $x$ as one-dimensional, but the following arguments also hold when $x$ is multidimensional.

In this case, the estimator $\hat p$ given by \eq{phat_w} is not unbiased in finite samples, but still unbiased in the asymptotic limit (consistent). We again compute the expectation of \eq{phat_w_alt} and take the expectation over the Bernoulli distributions $\probb(1, p(x_\ell))$ first, to obtain
\begin{equation}
    \Ex{\hat p} = \Ex{\frac{\sum_\ell^{\hat n} w(x_\ell) p(x_\ell)}{\sum_\ell^{\hat n} w(x_\ell)}}.
\end{equation}
Since the numerator and the denominator both contain the randomly sampled $x_\ell$, we cannot process this further without going to the asymptotic limit $\hat n \rightarrow \infty$, which allows us to replace the sums with expectations based on the law of large numbers,
\begin{equation}
    \Ex{\hat p} \xrightarrow{\hat n \to \infty}
    \frac{\Ex{\hat n} \, \Ex{w(x) \, p(x)}}{\Ex{\hat n} \, \Ex{w(x)}} = \bar p.
\end{equation}
This shows indirectly, that \eq{phat_w} is biased when $\hat n$ is finite.

\fig{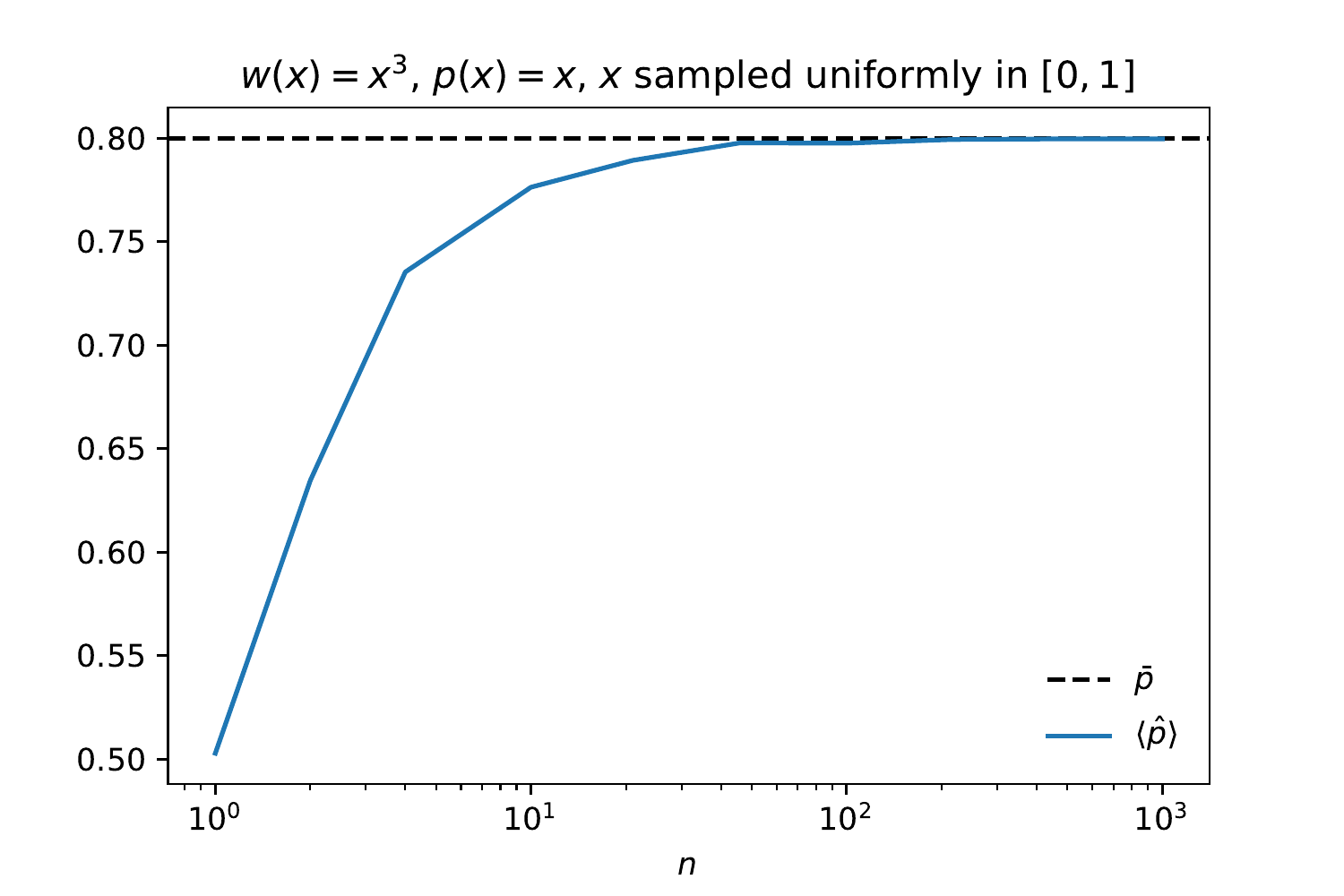}{Average of the estimator $\hat p$ (solid line) over 1000 independent trials for random $x$ drawn from a uniform distribution in the interval $[0, 1]$, efficiency $p(x) = x$ and weight $w(x) = x^3$. The dashed line indicates the asymptotic value $\bar p$.}

The bias is demonstrated in a numerical example, shown in \fg{weight_bias}. The bias can be substantial in small samples, if the weight distribution is very far from uniform. In that case, the bias should be estimated and subtracted.

The variance of $\hat p$ now differs from \eq{var_w_approx},
\begin{align}
    \Ex{(\hat p - \bar p)^2}
     & = \Ex{\left(\frac{\sum_\ell^{\hat n} X_\ell w_\ell}{\sum_\ell^{\hat n} w_\ell} - \bar p\right)^2}
    = \Ex{\frac{\sum_\ell^{\hat n} (X_\ell - \bar p)^2 w^2_\ell}{(\sum_\ell^{\hat n} w_\ell)^2}} \nonumber                                                                            \\
     & = \Ex{\frac{\sum_\ell^{\hat n} ((X_\ell - p_\ell) + (p_\ell - \bar p))^2 w^2_\ell}{(\sum_\ell^{\hat n} w_\ell)^2}} \nonumber                                                   \\
     & = \Ex{\frac{\sum_\ell^{\hat n} \big((X_\ell - p_\ell)^2 + (p_\ell - \bar p)^2 + 2 (X_\ell - p_\ell) (p_\ell - \bar p)\big) w^2_\ell}{(\sum_\ell^{\hat n} w_\ell)^2}} \nonumber \\
     & \xrightarrow{\hat n \to \infty} \frac{\Ex{\hat n} \, \Ex{\big(p(x) (1 - p(x)) + (p(x) - \bar p)^2\big) \, w^2(x)} }{\Ex{\hat n}^2 \, \Ex{w(x)}^2} \nonumber                    \\
     & = \frac{\Ex{p(x) (1 - p(x)) \, w^2(x)}}{n \, \Ex{w(x)}^2} + \frac{\Ex{(p(x) - \bar p)^2 \, w^2(x)} }{n \, \Ex{w(x)}^2}. \eql{var_w_approx_x}
\end{align}
As before, this formula is only asymptotically valid. If either $p(x)$ or $w(x)$ are constant, \eq{var_w_approx_x} reduces to \eq{var_w_approx}. Otherwise, the result computed from \eq{var_w_approx_x} can be larger or smaller than the result from \eq{var_w_approx}, depending on the interplay of $w(x)$ and $p(x)$.

An estimate for \eq{var_w_approx_x} can be constructed by computing estimates $\hat p_i$ and $\hat w_i$ over small intervals $\Delta x_i$, followed by replacing the expectations with arithmetic means over these intervals. A caveat of this approach is that one has to be aware of all variables on which both of $p$ and $w$ depend on. A simpler and more generic solution is to compute the variance of the estimator given by \eq{phat_w} with the bootstrap method \cite{Efron:1986hys}, which works in this case and all previous cases. There is no binning necessary then and all potential dependencies between $p$ and $w$ are taken into account.

\fig{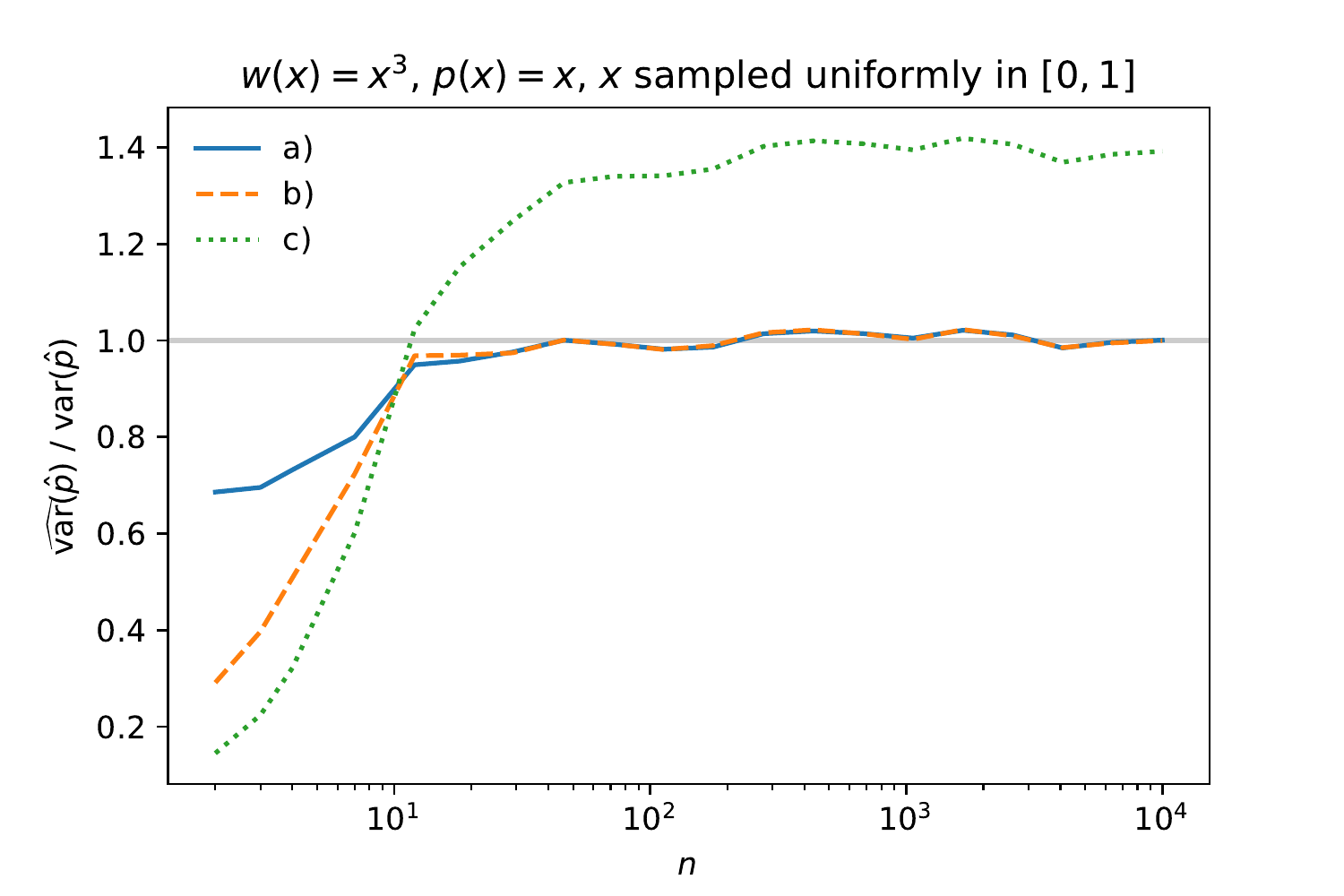}{Average variance estimate of $\hat p$ divided by the variance from Monte-Carlo simulation, using the same setting as for \fg{weight_bias}. The variance estimates are computed with a) a finite sample approximation of \eq{var_w_approx_x} as described in the text, b) the bootstrap method, c) \eq{var_w_approx} which is not applicable to this case.}

Variance estimates computed with the binned approach and the bootstrap are compared with each other in \fg{weight_variance}, and with the wrong estimate from \eq{var_w_approx}. The first two estimates converge to the right value rather rapidly, while \eq{var_w_approx} is completely off since it is not applicable to this case.

\section{Counts with additional fluctuations}

We now consider the common case where the estimates $\hat n_k$ cannot be obtained by direct counting, but need to be estimated in some other way that inflates their variance. A typical case in high-energy physics is that the estimates $\hat n_k$ are obtained from a fit to the invariant mass distribution of decay candidates. Such a fit removes the average background contribution, but it adds additional variance to the estimate, since the background fluctuates.

This case and any other, where an additional random contribution is added to a sample from a Poisson process, can be modeled in the following way,
\begin{equation}
    \hat n_1 = n_1 + z_1 \sqrt{n_1} + z_2 \, \sigma_{1,b},
\end{equation}
and likewise for $\hat z_2$. Here, the $z$ are random numbers with $\Ex{z} = 0$ and $\Ex{z^2} = 1$ and $\sigma_{k,b}^2$ is the additional variance that does not originate from the Poisson process. We assume that the random process which generates the extra variance is uncorrelated to the Poisson process. The variance of $\hat n_k$ then is
\begin{equation}
    \var(\hat n_k) = n_k + \sigma_{k,b}^2.
\end{equation}
In general, $\sigma^2_{k,b}$ is difficult to estimate, but an estimate for $\var(\hat n_k)$ is usually readily available. In the fit that was just described, $\var(\hat n_k)$ is estimated by inverting the Hessian matrix of the log-likelihood at the best fit values and taking the diagonal component that corresponds to the parameter $\hat n_k$ (the so-called HESSE method in MINUIT \cite{James:1975dr}). We thus use the substitution
\begin{equation}
    \sigma_{k,b} = \sqrt{\var(\hat n_k) - n_k}.
    \eql{s1b_subs}
\end{equation}

We compute the variance of the efficiency estimate $\hat p$ under these conditions. The estimate is \begin{equation}
    \hat p = \frac{\hat n_1}{\hat n_1 + \hat n_2} = \frac{n_1 + z_1 \sqrt{n_1} + z_3\, \sigma_{1,b}}{n_1 + z_1 \sqrt{n_1} + z_3\, \sigma_{1,b} + n_2 + z_2 \sqrt{n_2} + z_4\, \sigma_{2,b}},
\end{equation}
and its variance is to second order
\begin{equation}
    \var(\hat p) \approx J C_z J^T,
\end{equation}
with the row-matrix $J$ given by the derivatives $J_i = \partial \hat p / \partial z_k$ evaluated at $z_k = 0$ and the covariance matrix
\begin{equation}
    C_z = \begin{pmatrix}
        1 & 0 & 0    & 0    \\
        0 & 1 & 0    & 0    \\
        0 & 0 & 1    & \rho \\
        0 & 0 & \rho & 1    \\
    \end{pmatrix},
\end{equation}
for the $z_k$, where $\rho$ is the correlation coefficient  between the two background contributions. Estimating $\rho$ from data is not straight-forward, but fortunately $\rho$ is often zero. If $n_1$ and $n_2$ are estimated from independent samples, the backgrounds are independent, too. If the estimation of the backgrounds does not use shared parameters, the background estimates are independent as well. We obtain
\begin{equation}
    \var(\hat p) \approx \frac{p (1 - p)} n + \frac{p^2 \sigma_{2,b}^2 + (1-p)^2 \sigma_{1,b}^2}{n^2}
    - 2 \rho \frac{p (1 - p) \sigma_{1,b} \sigma_{2,b}}{n^2}.
    \eql{var_extra_approx_raw}\end{equation}
which reduces to \eq{var_binom} for $\sigma_{k,b} = 0$. Substitution of $\sigma_{k,b}$ gives
\begin{equation}
    \var(\hat p) \approx \frac{n_1^2 \var(\hat n_2) + n_2^2 \var(\hat n_1)}{(n_1 + n_2)^4} - 2 \rho \frac{n_1 n_2 \sqrt{\var(\hat n_1) - n_1} \sqrt{\var(\hat n_2) - n_2}}{(n_1 + n_2)^4} .
    \eql{var_extra_approx}
\end{equation}
For $\rho = 0$ we get the simple formula
\begin{equation}
    \var(\hat p) \approx \frac{n_1^2 \var(\hat n_2) + n_2^2 \var(\hat n_1)}{(n_1 + n_2)^4}.
    \eql{var_extra_approx_simple}
\end{equation}

\eq{var_extra_approx} is a good approximation for $n \gg 1$ and $\sigma_{k,b} \ll n$. In contrast to the previous section, we will not attempt to compute $\var(\hat p)$ more accurately, since that would require additional information about the process which introduces the extra fluctuations. One needs the higher moments of $z_3$ and $z_4$, which are usually not readily available. It is possible to improve the result in the limit $\sigma_{k,b}/n \to 0$ by replacing the first term in \eq{var_extra_approx_raw} with \eq{var_poisson} and using \eq{fn_approx} for $f(n)$.

To demonstrate the validity of \eq{var_extra_approx_simple}, we simulate an analysis in which the estimates $\hat n_k$ are obtained from a fit to data which consists of a normal signal and uniformly distributed background. The peak is modelled with a normal distribution whose amplitude, location, and width are free parameters. The background is modelled with a second degree Bernstein polynomial. The simulation is run 1000 times. In each run, the number of successes and failures are sampled independently and drawn from Poisson distributions with expectations $p n$ and $(1-p)n$, respectively. The expected number of background events for both successes and failures is taken to be $0.2\, n$, and also Poisson distributed. Since the background events are independently sampled from the successes and failures, we have $\rho = 0$.

\fig{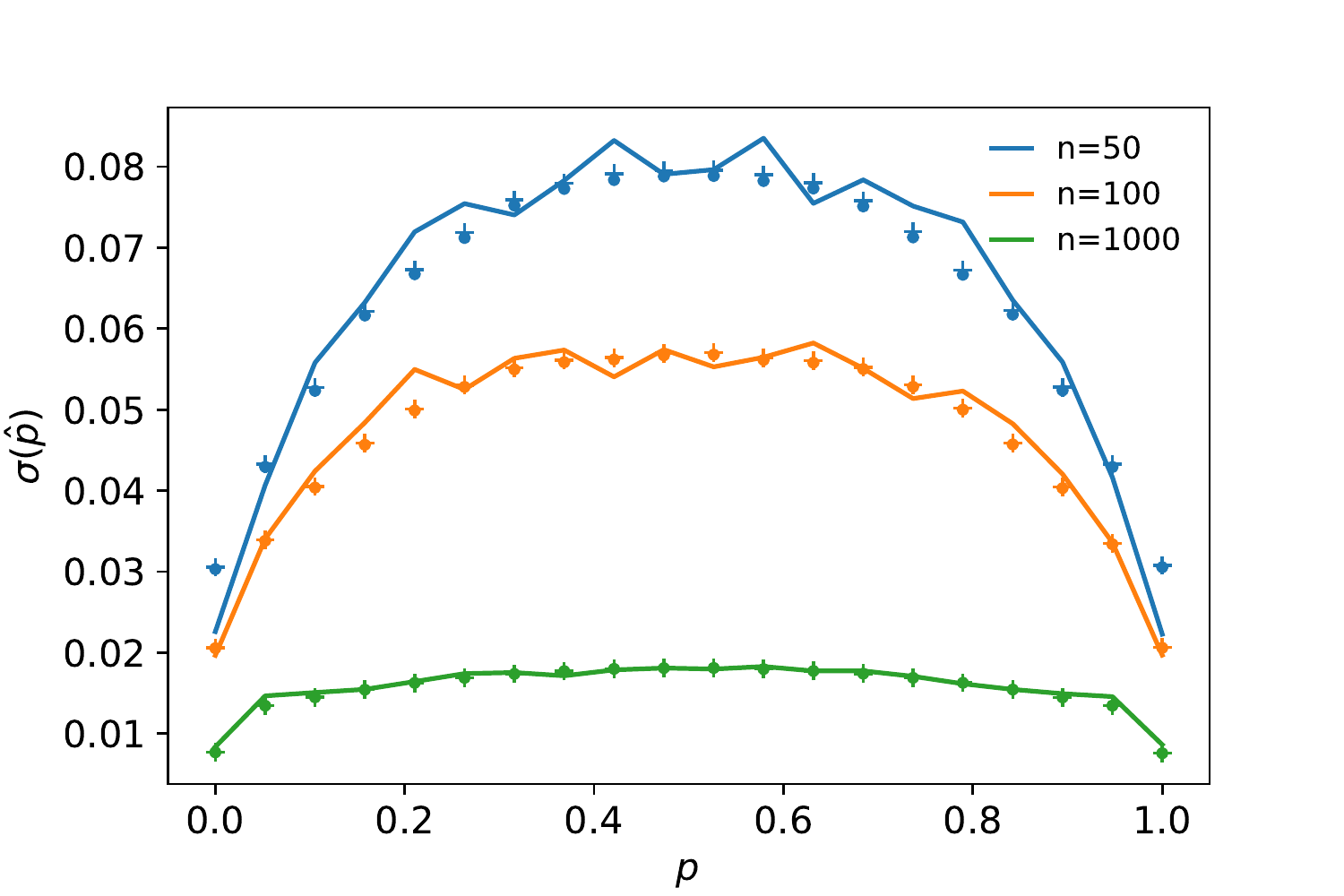}{Standard deviation of the efficiency estimate $\hat p$ for the simulated analysis described in the text, in which the estimates $\hat n_k$ are obtained from a fit to a mixture of signal and background. Shown are the standard deviations over the repetitions (lines) and the square-root of the average variance estimates (dots) for different total expected number of events $n$. An alternative calculation in which the first term in \eq{var_extra_approx_raw} is corrected with \eq{fn_approx} is also shown (crosses).}

The results of the simulation are shown in \fg{variance_extra}. The results of \eq{var_extra_approx_simple} show satisfactory agreement with the true variance of the estimate $\hat p$, but we also point out deviations up to 18\,\% (6\,\%) in case of $n=50$ ($n=1000$). Convergence is rather slow. The approximate formula underestimates the true variance for central values of $p$ and overestimates for values close to zero or one. This accuracy is usually good enough to draw an error bar in a plot, however. For more accurate calculations, bootstrap methods could be used \cite{Efron:1986hys}.

Sometimes, it is more convenient to estimate $\hat n_1$ and $\hat n$ instead of $\hat n_1$ and $\hat n_2$ to compute $\hat p$, so we also compute the variance for this case. The variance of $\hat n$ is $\var(\hat n) = n + \sigma^2_{1,b} + \sigma^2_{2,b} + 2 \rho\, \sigma_{1,b} \, \sigma^2_{2,b}$ and the substitution for $\sigma_{2,b}$ becomes
\begin{equation}
    \sigma_{2,b} = \sqrt{\var(\hat n) - \var(\hat n_1) - n + n_1 (1 - \rho^2) + \rho^2 \var(\hat n_1)} -\rho \sqrt{\var(\hat n_1) - n_1},
\end{equation}
which yields
\begin{multline}
    \var(\hat p) \approx \frac{n^2 \var(\hat n_1) + \var(\hat n) n_1^2 - 2 n n_1 \var(\hat n_1) }{n^4}
    + 2 \rho\frac{n_1}{n^3} \Bigg(
    \rho (\var(\hat n_1) - n_1) - \\
    \sqrt{\var(\hat n_1) - n_1}
    \sqrt{
            \rho^2 \big(\var(\hat n_1) - n_1\big) + \var(\hat n) - n + n_1 - \var(\hat n_1)
        }
    \Bigg).
\end{multline}
This formula is rather complicated, but when $\rho$ is zero, it reduces to
\begin{equation}
    \var(\hat p) \approx \frac{n^2 \var(\hat n_1) + \var(\hat n) n_1^2 - 2 n n_1 \var(\hat n_1) }{n^4}.
    \eql{var_extra_n_n1_simple}
\end{equation}
We note that the estimate of this variance computed from sample estimates $\hat n$, $\hat n_1$ $\widehat\var(\hat n)$, and $\widehat\var(\hat n_1)$ can become negative, due to random fluctuations in those estimates. This is not a defect of the formula itself, the true variance $\var(\hat p)$ is never negative, since $\var(\hat n
    ) > n$, $\var(\hat n_1) > n_1$, and $n > n_1$.

\section{Generalized Wilson intervals}

The ansatz of the standard Wilson interval is to regard the normalized residual of $\hat p$ as approximately standard-normal distributed, so that its square approximately satisfies,
\begin{equation}
    z^2 \approx \frac{(\hat p - p)^2}{\var(\hat p)},
    \eql{wilson_base}
\end{equation}
where $z$ is standard-normally distributed. For a given value of $z^2$ and the observed value $\hat p$, one finds the interval in $p$ that satisfies the equation. The value $z^2$ is computed from the inverse of the cumulative $\chi^2$ distribution with one degree of freedom, evaluated at the desired coverage probability. The crucial improvement of the Wilson interval over the normal approximation interval is that $\var(\hat p)$ is correctly considered as a function of $p$ when solving the equation, while the latter uses a constant value for $\var(\hat p)$, evaluated at the point $p = \hat p$.

We follow this approach to compute a Wilson interval for the case of Poisson-distributed counts without extra fluctuations. We insert \eq{var_poisson} in \eq{wilson_base} and get a quadratic equation in $p$,
\begin{equation}
    (\hat p - p)^2  = z^2 \frac{p(1 - p)}{n} f(n),
\end{equation}
which we solve for $p$ to obtain a Wilson interval
\begin{equation}
    p_{1,2} = \frac 1 {1 + \frac{z^2}n f(n)} \left(
    \hat p + \frac{z^2}{2n} f(n) \pm
    \frac{z}{n} \sqrt{\hat p (1 - \hat p) n f(n) + \frac{z^2}4 f(n)^2}
    \right),
    \eql{generalized_wilson_interval}
\end{equation}
To obtain the standard 68\,\% interval, one sets $z = 1$. One can insert \eq{fn_approx} for $f(n)$. We note that this is interval is not a full implementation of Wilson's approach, since deviations of $n$ from $\hat n$ are not taken into account. For small $n$, this interval may be too narrow, as we will see in the next section. Therefore, we do not recommend this Wilson interval over the standard one, even though it is computed with an improved estimate of the variance.

For weighted samples, the generalized Wilson interval is \eq{generalized_wilson_interval} with $n$ replaced by $n_\text{eff}$ and using \eq{fn_o4} to compute $f(n_\text{eff})$. In this case, the correction $f(n_\text{eff})$ is helpful to improve the coverage of the intervals in small samples.

We compute a generalized Wilson interval for counts with independent extra fluctuations (assuming $\rho = 0$). We insert \eq{var_extra_approx_raw} into \eq{wilson_base} and again get a quadratic equation in $p$,
\begin{equation}
    (\hat p - p)^2  = \frac{z^2}{n^2} \left((\sigb{1}^2 + \sigb{2}^2 - n) p^2 + (n - 2 \sigb{1}^2) p + \sigb{1}^2 \right).
\end{equation}
Solving for $p$ yields
\begin{multline}
    p_{1,2} = \frac{1}{1 + \frac{z^2}{n} \big(1 - \frac{\sigb1^2 + \sigb2^2}{n}\big)} \Bigg(
    \hat p + \frac{z^2}{2 n} \bigg(1 - 2\frac{\sigb1^2}{n}\bigg) \; \pm \\
    \frac{z}{n} \sqrt{
            \hat p^2 (\sigb1^2 + \sigb2^2 - n) + \hat p (n - 2 \sigb1^2) + \sigb1^2 + \frac{z^2}{4} \bigg(1 - 4 \frac{\sigb1^2 \sigb2^2}{n^2}\Big) \bigg)
        }
    \Bigg). \eql{wilson_extra}
\end{multline}
The $\sigb{k}$ are then substituted according to \eq{s1b_subs}.

\fig{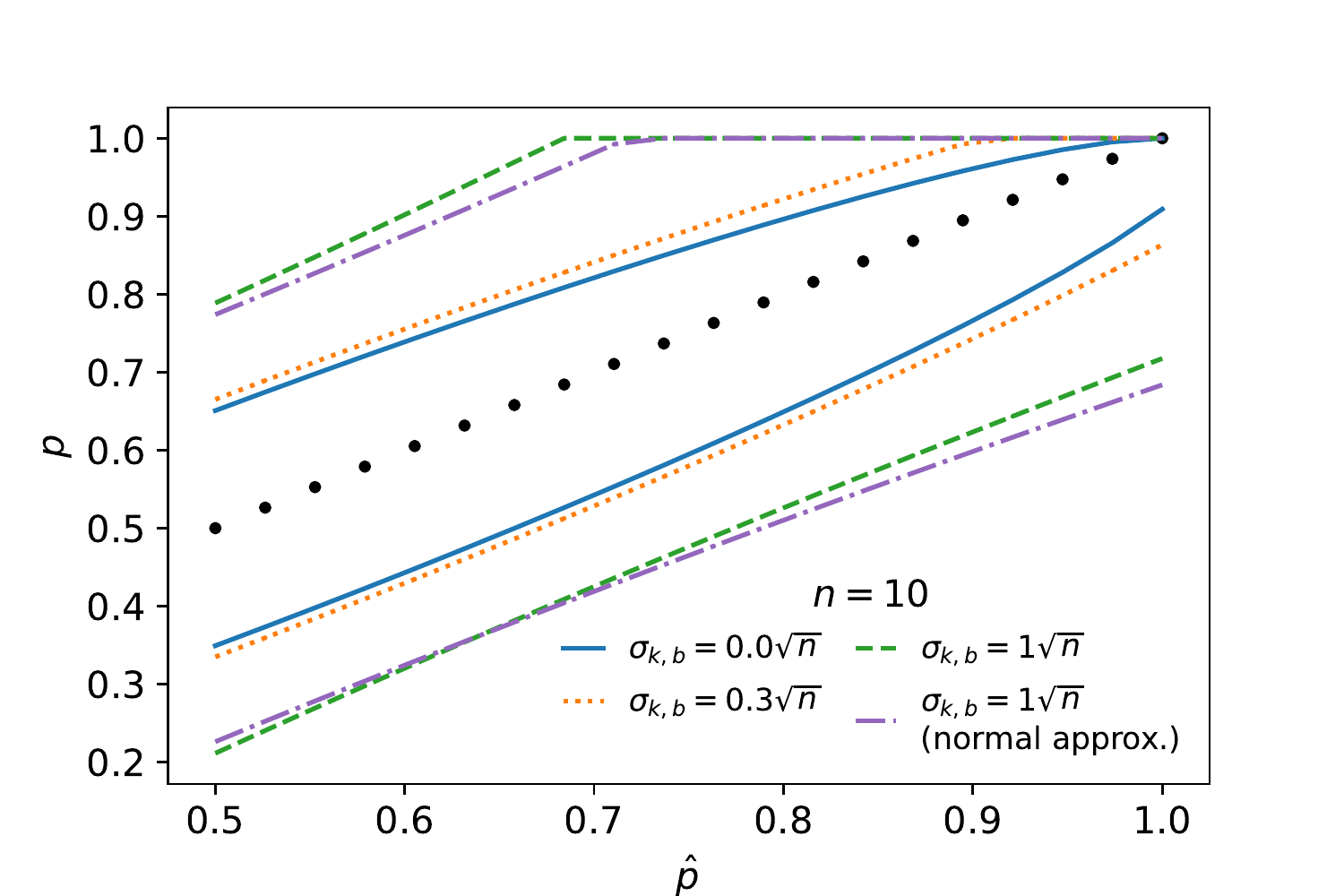}{Wilson intervals computed with \eq{wilson_extra}, when extra fluctuations are present. Shown are intervals as a function of the estimate $\hat p$ for $n = 10$ and different amounts of additional fluctuations. Also shown for comparison is the normal approximation interval $\hat p \pm \sqrt{\var(\hat p)}$ for one of the cases. All intervals have been clipped to the range $[0, 1]$.}

The generalized Wilson interval reduces to the original for $\sigma_{k,b} \to 0$. Results for non-zero values of $\sigma_{k,b}$ are shown in \fg{wilson_extra}. When the extra fluctuations are large, the interval becomes similar to the normal approximation interval $\hat p \pm \sqrt{\var(\hat p)}$, but it has a better limit for $\sigma_{k,b} \to 0$. The original Wilson interval has the nice property that the interval boundaries are automatically restricted to the range $p \in [0, 1]$, but this is not the case for the generalized interval.

\section{Coverage probability of different intervals}

The coverage probability is given by the fraction of intervals that cover the true value in repeated identical random trails. We compute the coverage probability of the Wilson interval and other popular intervals:
\begin{itemize}
    \item Clopper-Pearson interval \cite{10.1093/biomet/26.4.404}. This is based on the exact Neyman construction of the interval. It is guaranteed to cover the true value with at least the requested coverage probability.
    \item Normal approximation interval. Like the Wilson interval, this interval assumes that sample deviations from the true value are normally distributed, but the variance of these deviations is considered to be constant, with $\var(\hat p)$ evaluated at $p = \hat p$.
    \item Bayesian intervals. We calculate Bayesian credible intervals for a uniform prior in $p$ and the binomial Jeffreys prior \cite{doi:10.1098/rspa.1946.0056}, respectively. Bayesian intervals are not designed to have any particular coverage probability, but can have good coverage properties in practice.
\end{itemize}

It should be noted that for simplicity we only consider the uniform and the binomial Jeffreys prior $J_B(p)$ in order to probe the sensitivity of the results with respect to the choice of the prior. The Jeffreys prior $J_P(p)$ for the Poisson case is
\begin{equation}
    J_P(p) = J_B(p)\times\left\{
    \begin{array}{l}
        1/(1-p) \;\mbox{for}\;  p\leq 1/2 \\
        1/p \;\;\;\;\;\;\;\;\;\;  \mbox{for}\;  p > 1/2
    \end{array}
    \right.
    \quad\mbox{with}\quad
    J_B(p) = \frac{1}{\sqrt{p(1-p)}} \;.
\end{equation}
It has the same singular behavior for $p\to 0$ and $p\to 1$ as the binomial prior, but is more uniform for intermediate values, and thus would interpolate between $J_B(p)$ and the uniform prior.

\begin{figure}
    \includegraphics[width=0.49\textwidth]{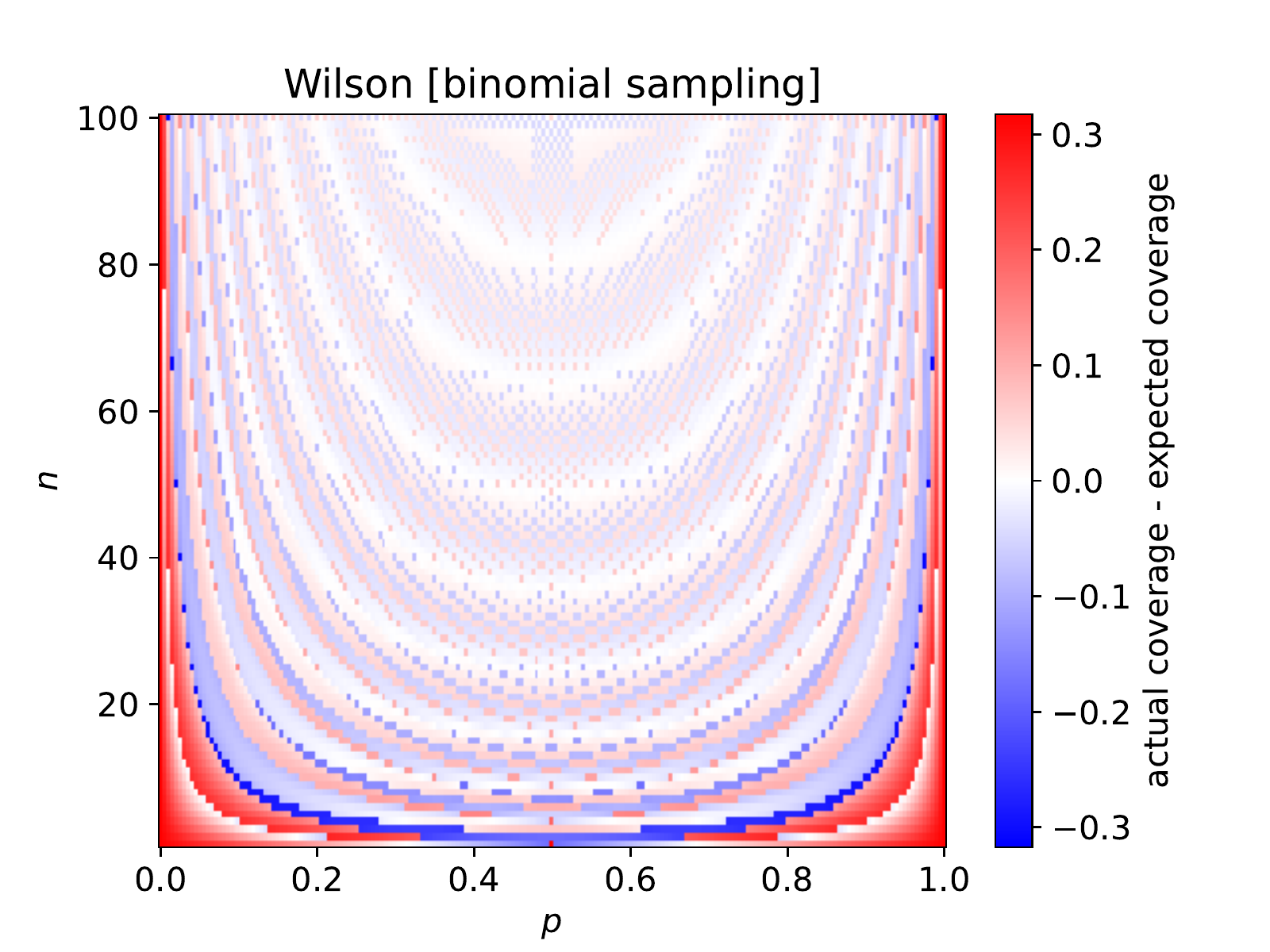}
    \includegraphics[width=0.49\textwidth]{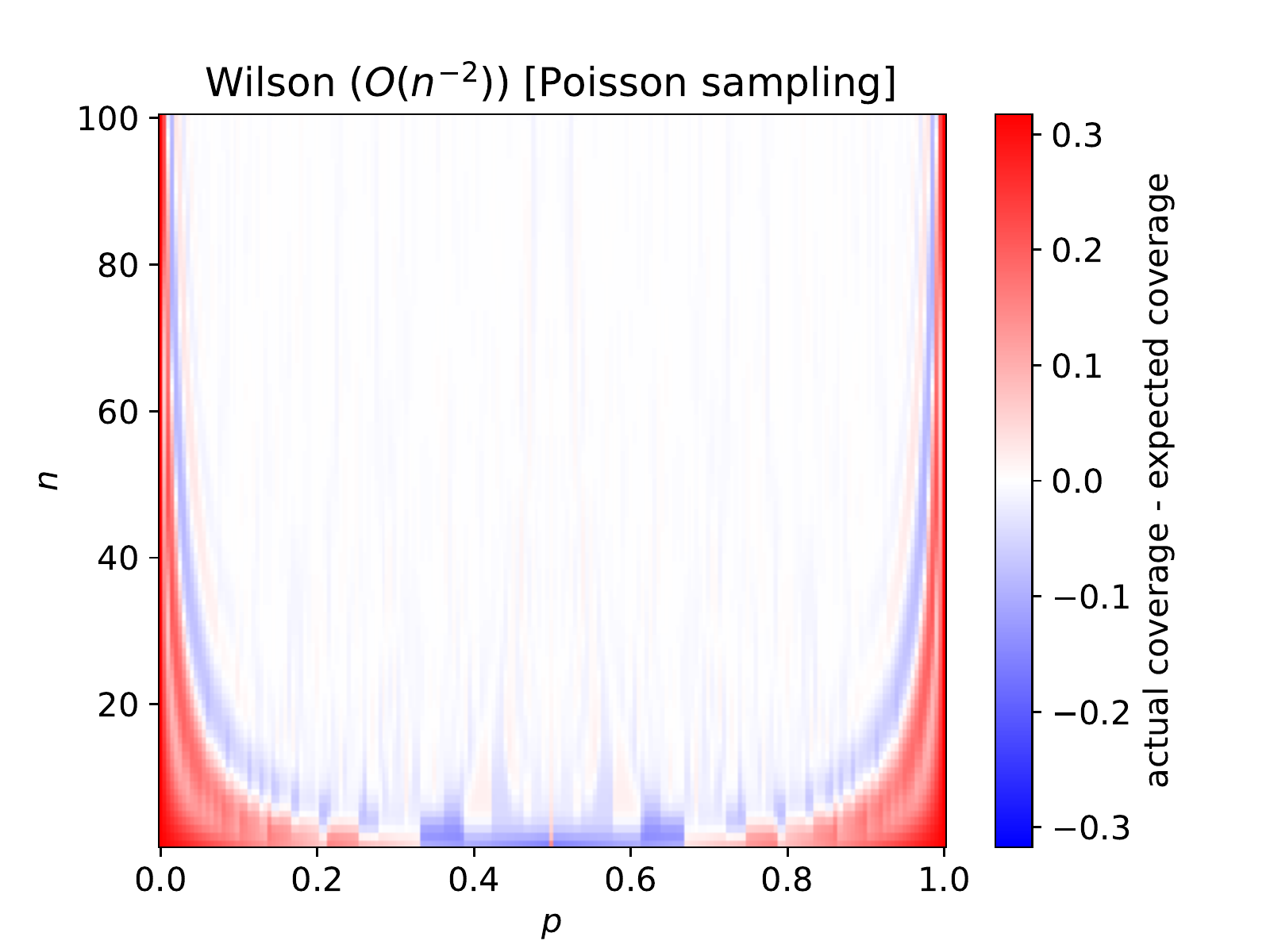}
    \includegraphics[width=0.49\textwidth]{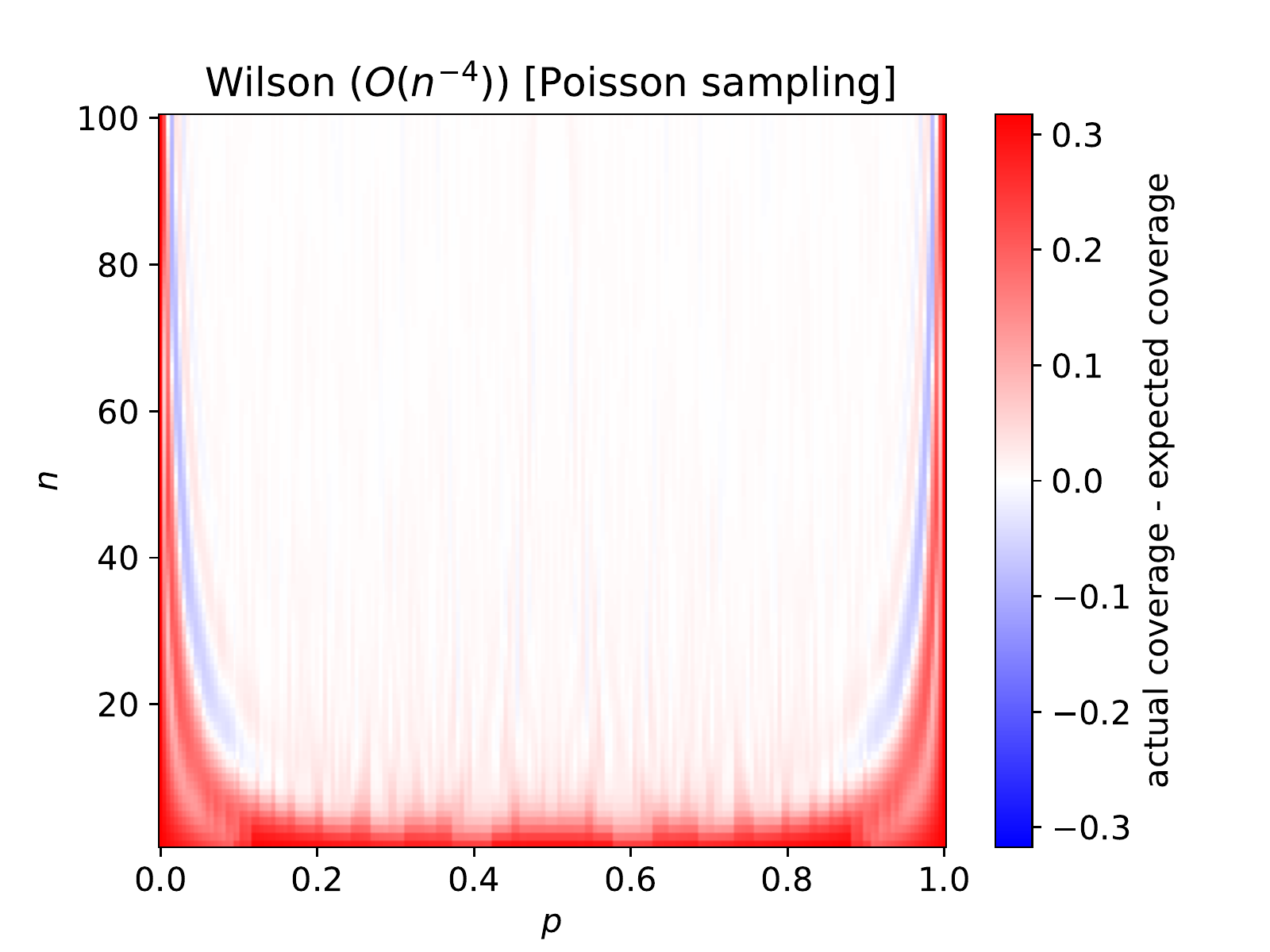}
    \includegraphics[width=0.49\textwidth]{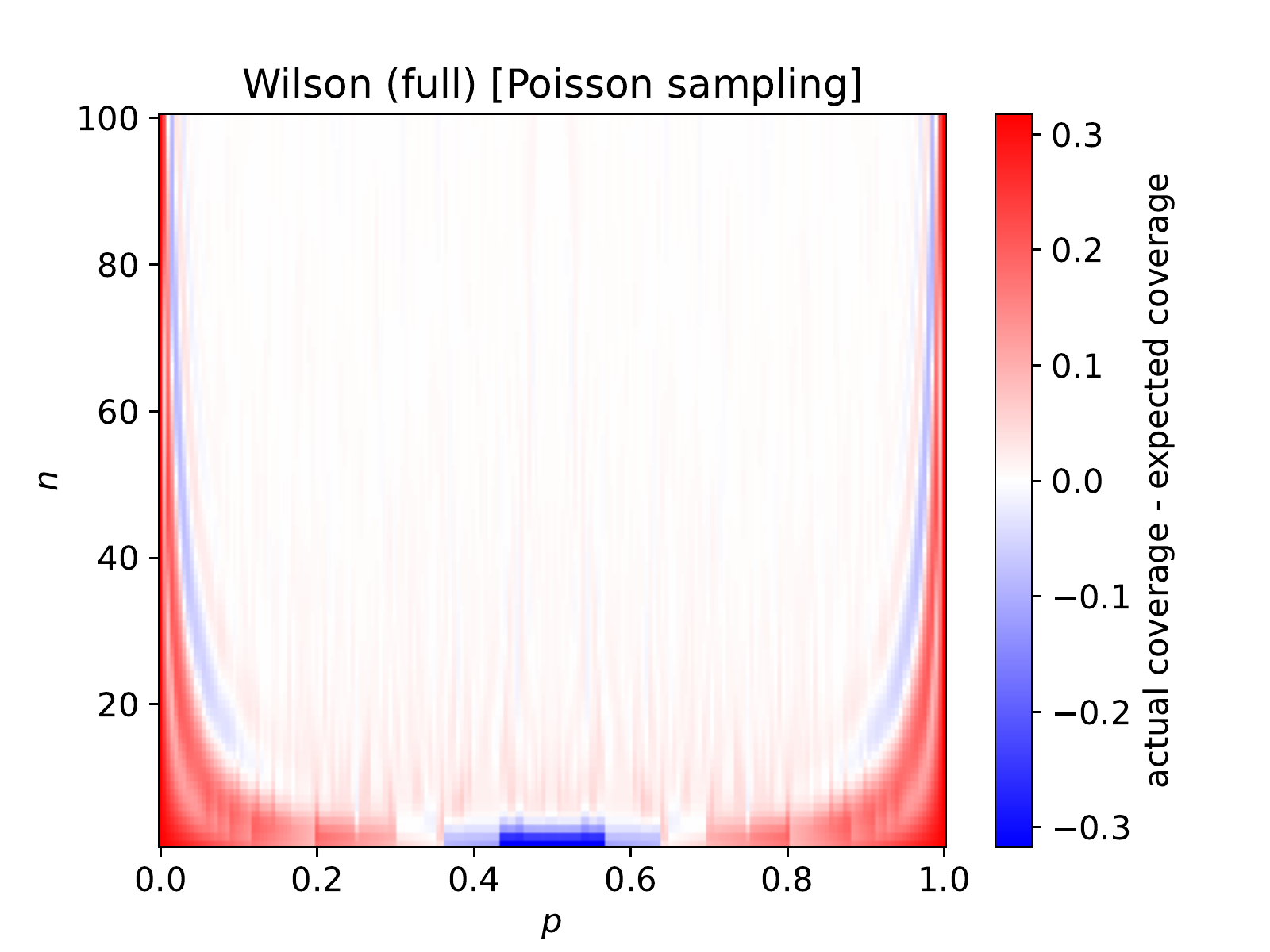}
    \caption{Deviation of the coverage probability from the expected value as a function of $p$ and the expected total number of events $n$. Samples in the top left plot are drawn from the binomial distribution, for the other plots they are drawn from the Poisson distribution. The plot in the top right corner shows the standard Wilson interval based on \eq{var_binom}. The plot in the bottom left corner shows the generalized Wilson interval based on \eq{fn_o4} and the plot in the bottom right corner shows the interval based on \eq{fn_approx}.}
    \label{fig:wilson_cov}
\end{figure}

\begin{figure}
    \includegraphics[width=0.49\textwidth]{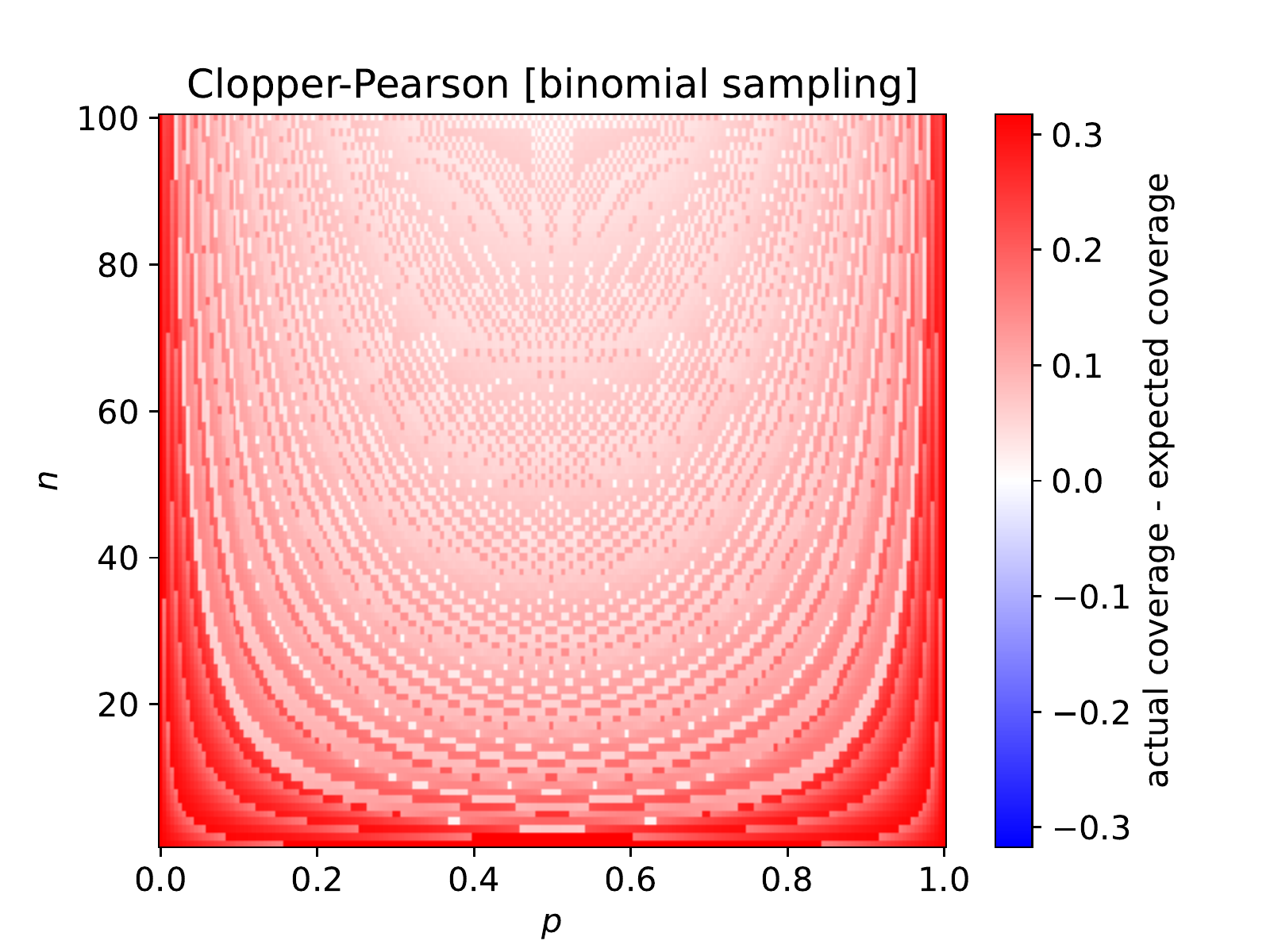}
    \includegraphics[width=0.49\textwidth]{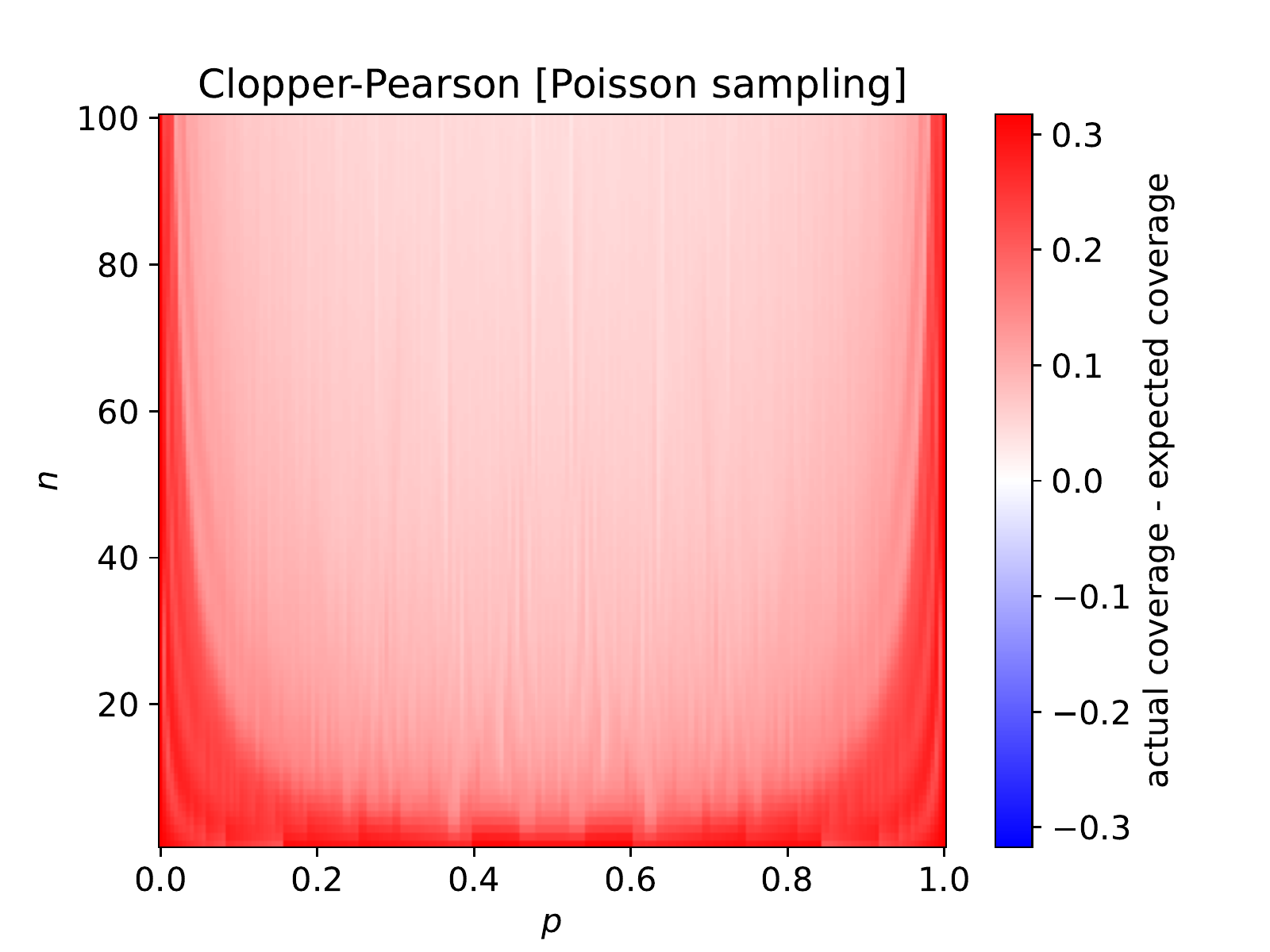}

    \includegraphics[width=0.49\textwidth]{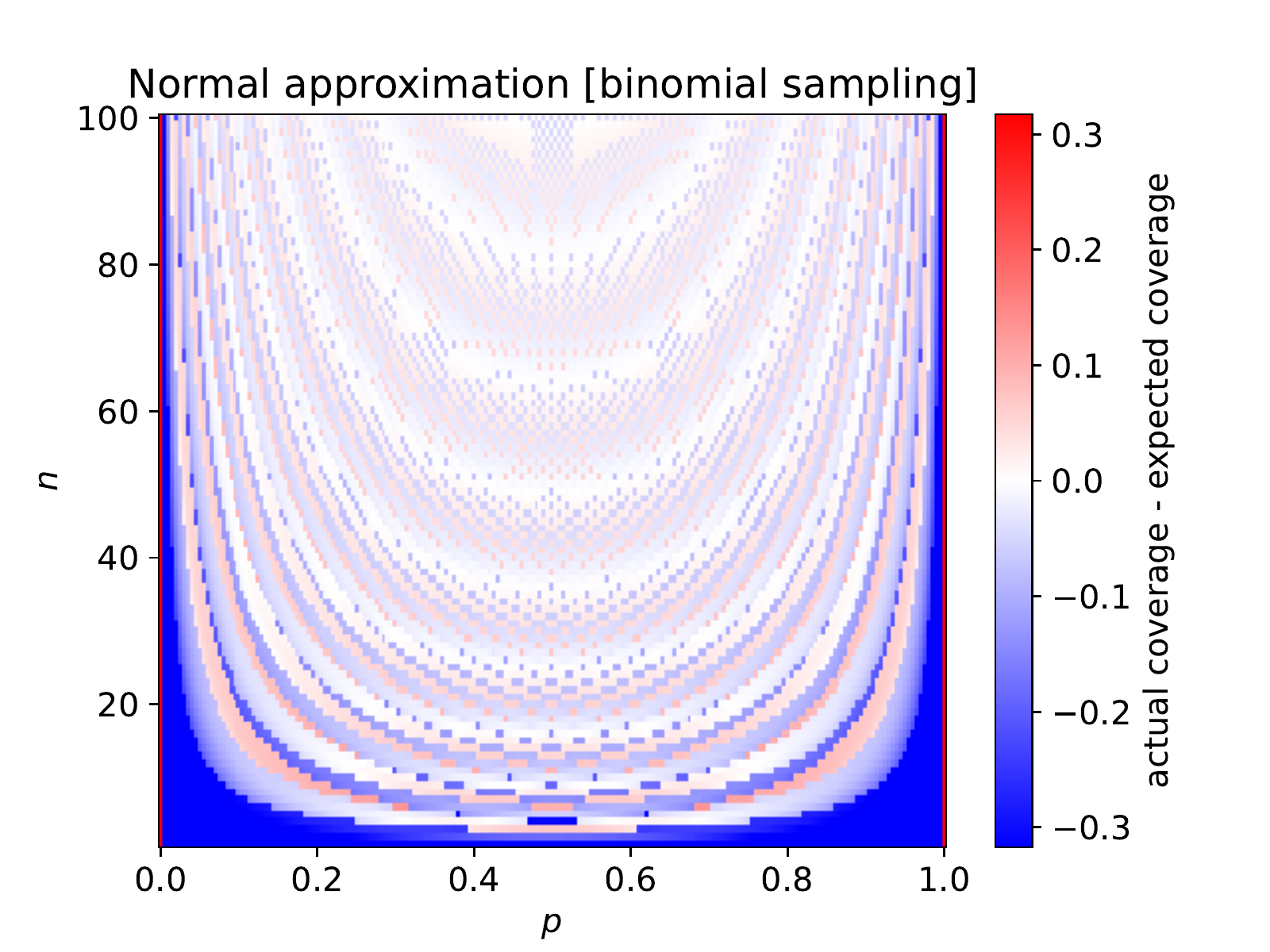}
    \includegraphics[width=0.49\textwidth]{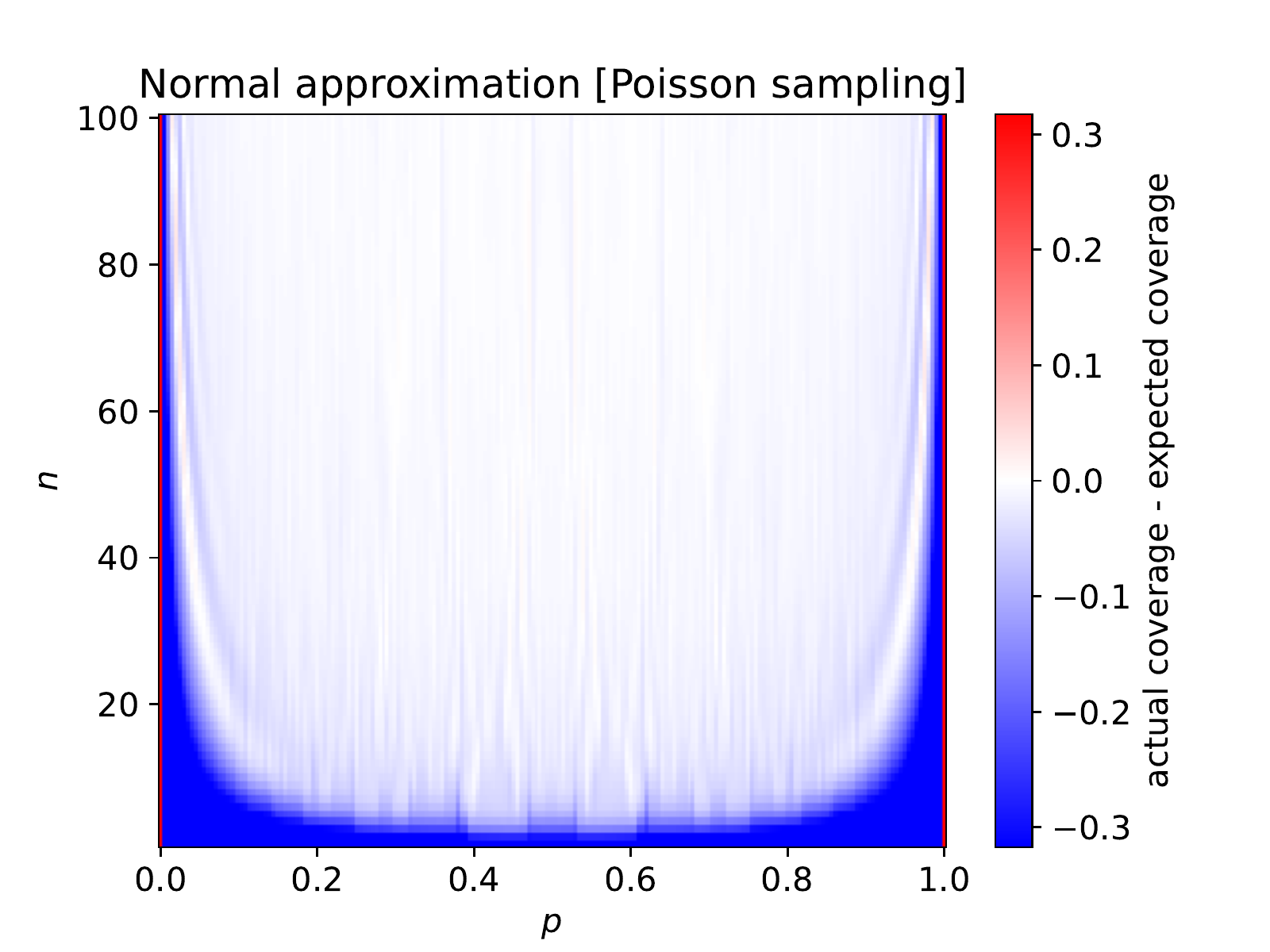}

    \includegraphics[width=0.49\textwidth]{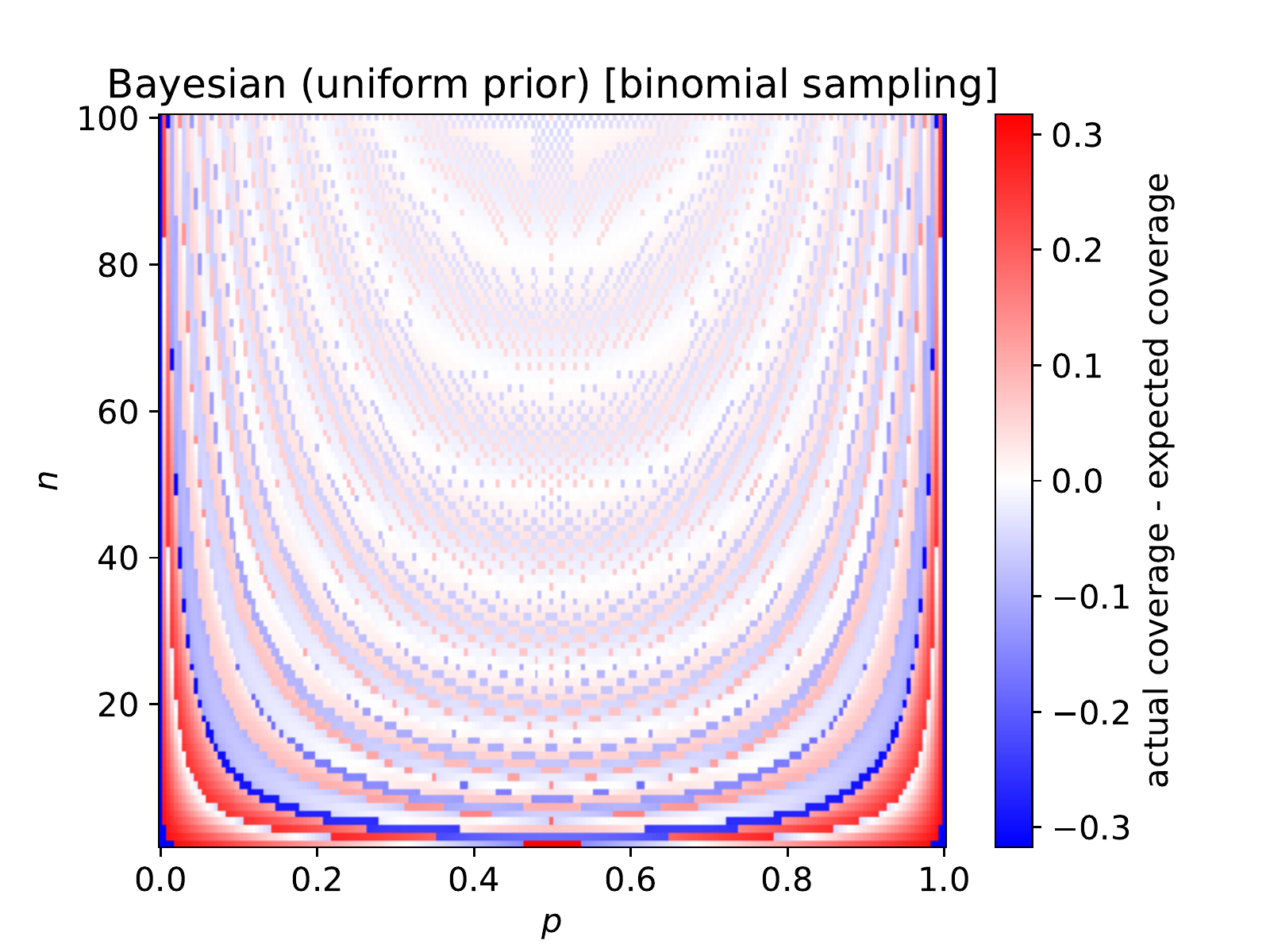}
    \includegraphics[width=0.49\textwidth]{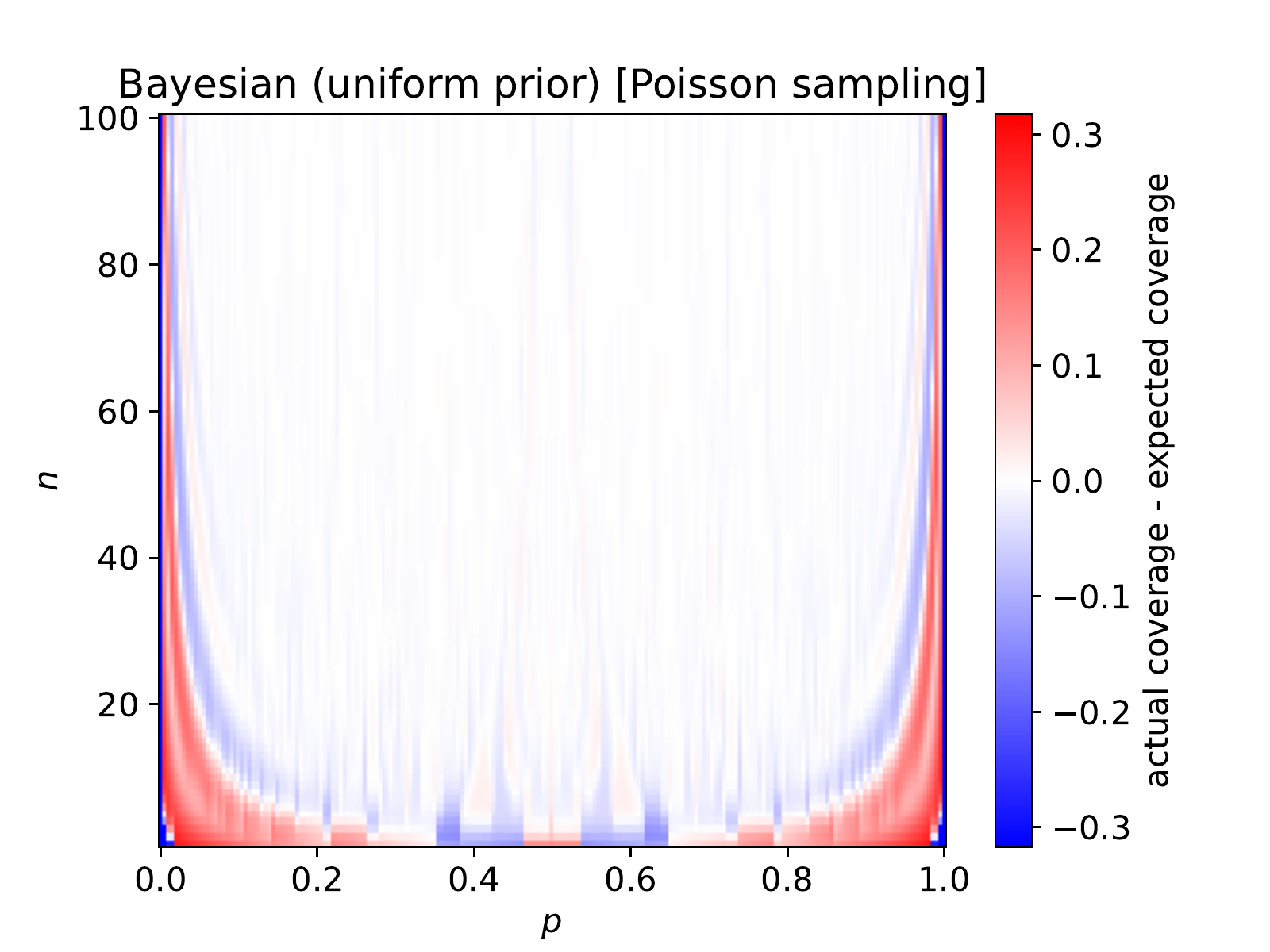}

    \includegraphics[width=0.49\textwidth]{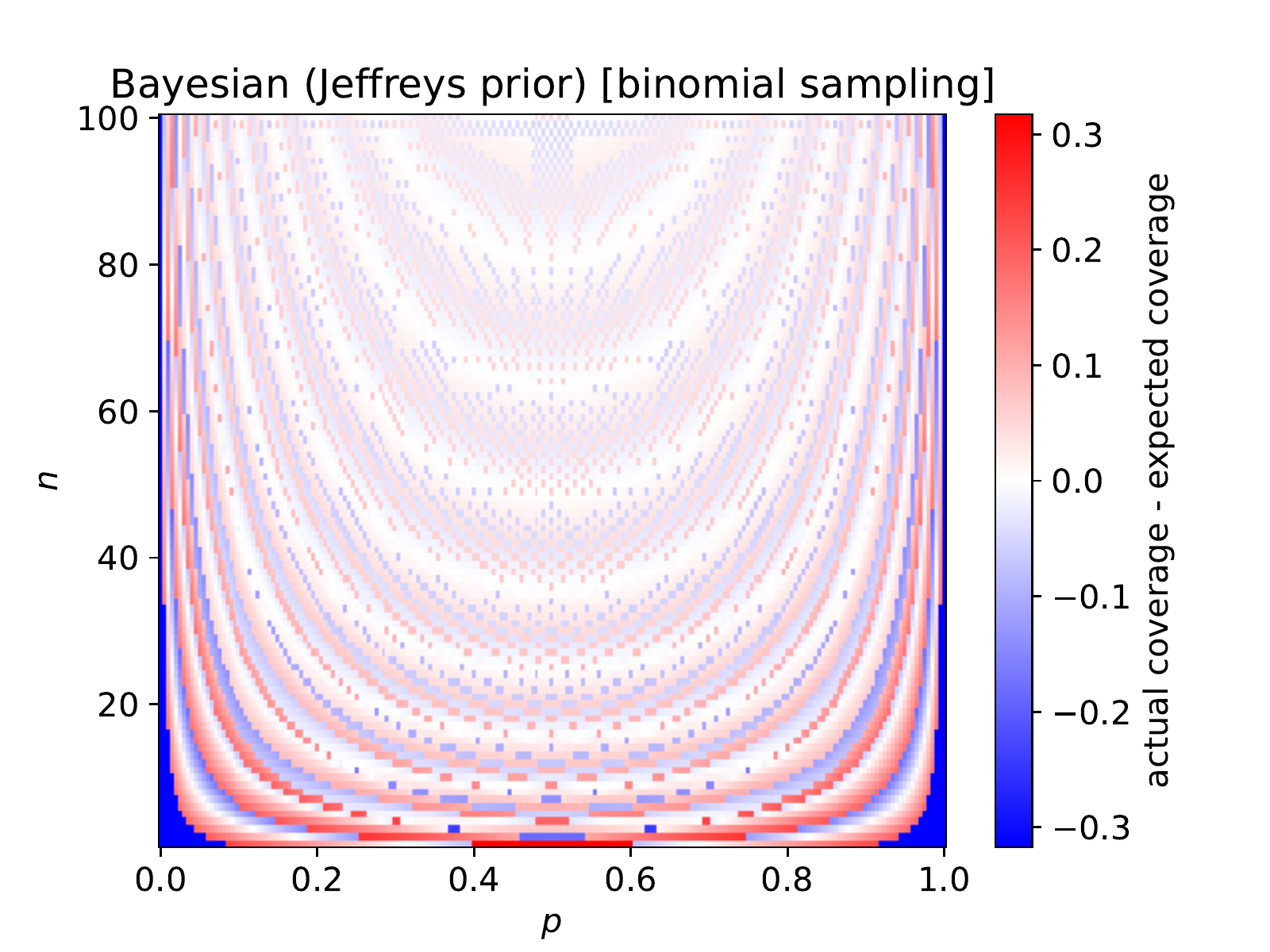}
    \includegraphics[width=0.49\textwidth]{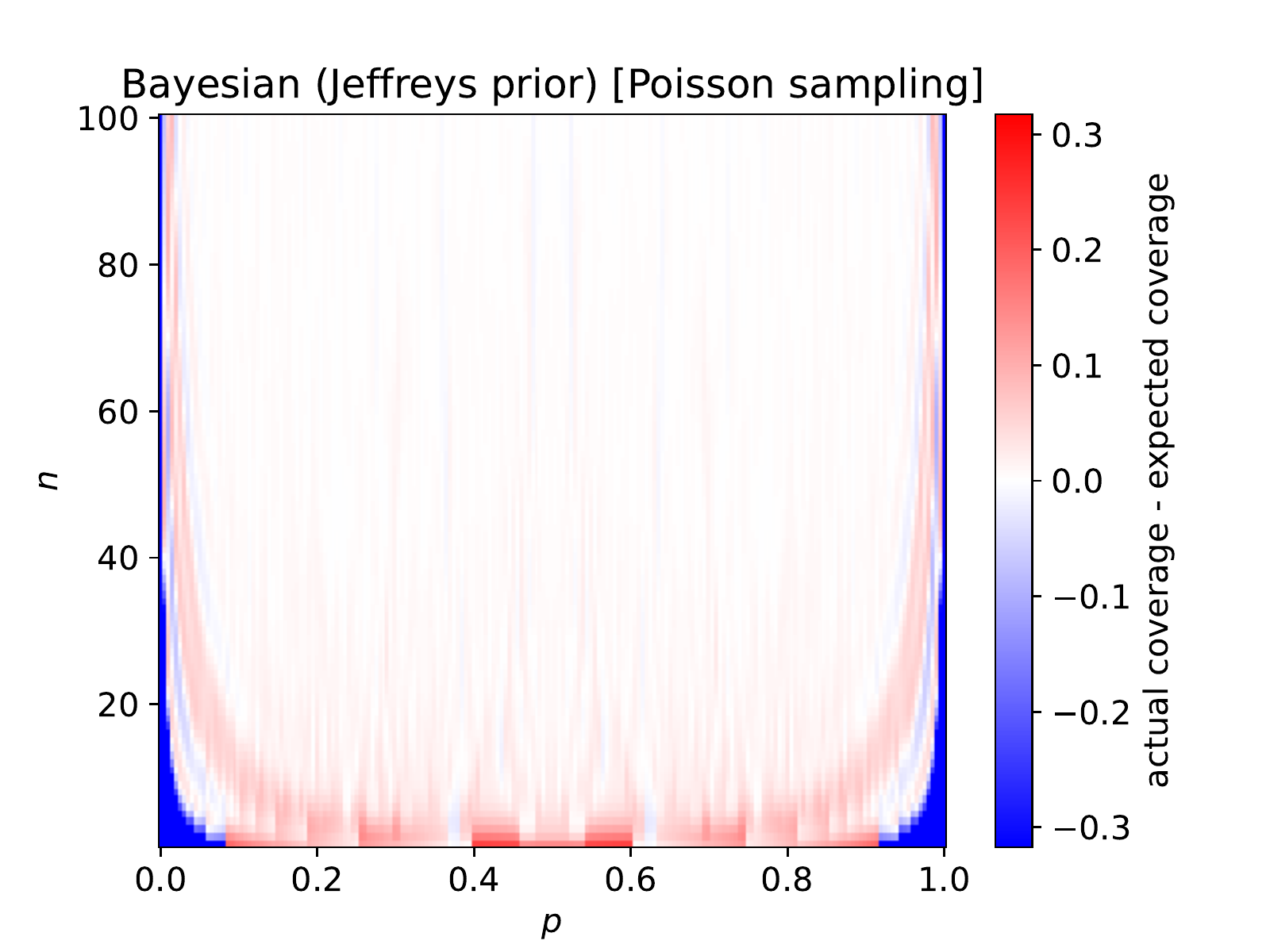}
    \caption{Deviation of the coverage probability from the expected value as a function of $p$ and the expected total number of events $n$. Samples are drawn from the binomial (Poisson) distribution for the plot on the left-hand (right-hand) side.}
    \label{fig:other_cov}
\end{figure}

\begin{figure}
    \includegraphics[width=\textwidth]{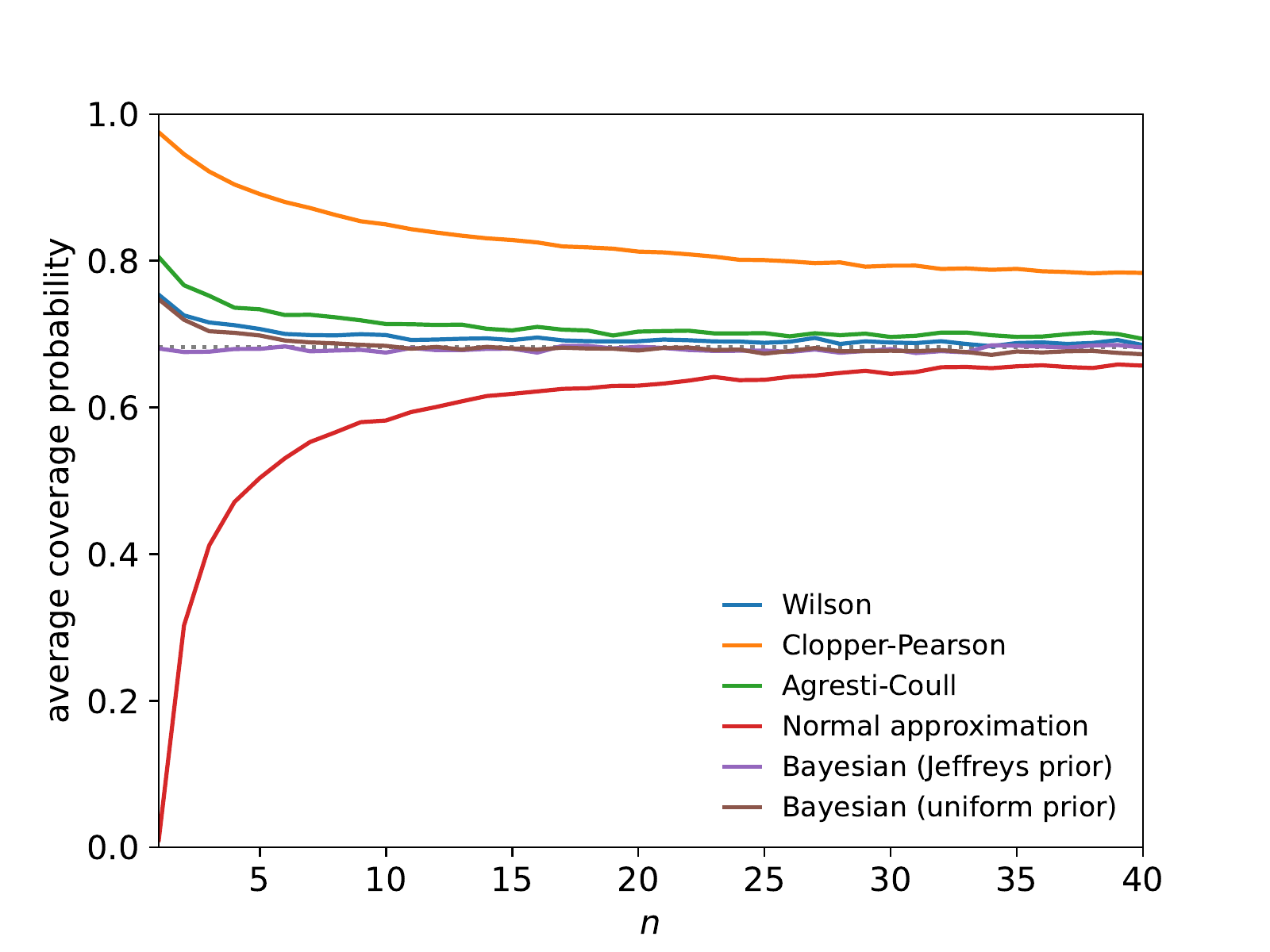}
    \includegraphics[width=\textwidth]{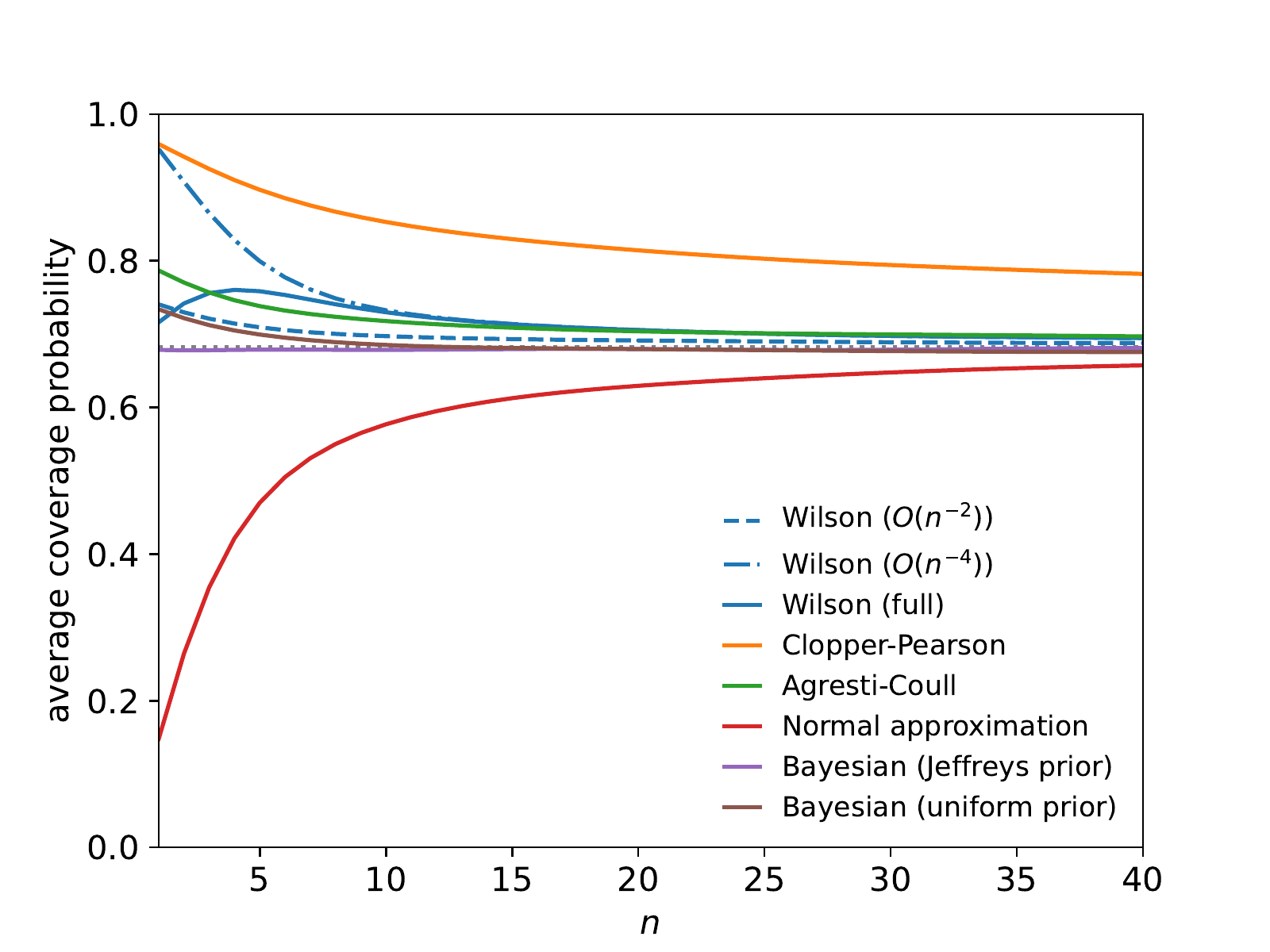}
    \caption{Coverage probability averaged over $p$ as a function of the expected total number of events $n$. Samples are drawn from the binomial (Poisson) distribution on in the upper (lower) plot.}
    \label{fig:average_cov}
\end{figure}

We compute the coverage probability of these intervals for binomially and Poisson distributed samples. The deviation of the actual coverage probability from the expected probability is shown as a function of the true efficiency $p$ and the expected total number of counts $n$ in \fg{wilson_cov} and \fg{other_cov}. The coverage probability is not a smooth function of $p$ and $n$ due to the discreteness of the binomial and Poisson distributions. This is very apparent in case of binomially distributed samples, and to a lesser degree also in case of Poisson distributed samples. The coverage probability averaged over values of $p$ is shown in \fg{average_cov}, using a uniform prior for $p$. The choice of this prior is arbitrary, but the qualitative results that we draw do not depend strongly on the prior.

We evaluate the results according to these criteria
\begin{itemize}
    \item Intervals with an average coverage probability closer to the expected value are preferred.
    \item Methods are disqualified if for some combinations of $p$ and $n$ the 68\% intervals have zero coverage.
          %\item Intervals are disqualified, if they do not cover the true value at all for some combinations of $p$ and $n$.
\end{itemize}
The second condition

The Clopper-Pearson and normal approximation intervals severely over- and undercover, respectively. When it is essential to be conservative, the Clopper-Pearson interval is the correct choice, but not in most practical cases. The average coverage probability is closest to the expected probability for the Bayesian credible interval with the Jeffreys prior. However, the Bayesian intervals have zero coverage for some values of $n$ and $p$, which violates a basic expectation one has for confidence intervals.

The Wilson intervals have a coverage probability close to the expected value and only slightly overcover on average. They never have zero coverage for some values of $n$ and $p$. The standard Wilson interval based on the variance for binomially distributed samples in \eq{var_binom} also performs well when the samples are Poisson distributed. It overcovers when $p$ is close to zero and one, and slightly undercovers when $p$ is near 1/2. When the Wilson interval is constructed with the correct variance for Poisson distributed samples, the latter is more pronounced. Undercoverage can be avoided completely by using the approximate variance based on \eq{fn_o4}, which makes the interval conservative for small $n$.

In conclusion, the standard Wilson interval is a good choice even when the samples are Poisson distributed. Its coverage probability is close to the expected value, it undercovers only mildly, and is readily available in statistical software or easily implemented by hand. In contrast to the other intervals, the Wilson interval can be easily generalized to situations where the samples are weighted and where the counts have additional fluctuations, since one merely has to replace the variance in the derivation of the interval.

\section*{Acknowledgements}

HD acknowledges funding by the Deutsche Forschungsgemeinschaft (DFG, German Research Foundation) -- project no. 449728698. The calculations in this paper use the following scientific software libraries: Numba \cite{siu_kwan_lam_2021_5524874}, ROOT \cite{Brun:1997pa}, Numpy \cite{Harris:2020xlr}, SciPy \cite{Virtanen:2019joe}, iminuit \cite{Dembinski:2020a}, matplotlib \cite{Hunter:2007ouj}, and SymPy \cite{Meurer:2017yhf}.

\bibliographystyle{model1-num-names}
\bibliography{main}

\begin{thebibliography}{13}
\expandafter\ifx\csname natexlab\endcsname\relax\def\natexlab#1{#1}\fi
\providecommand{\url}[1]{\texttt{#1}}
\providecommand{\href}[2]{#2}
\providecommand{\path}[1]{#1}
\providecommand{\DOIprefix}{doi:}
\providecommand{\ArXivprefix}{arXiv:}
\providecommand{\URLprefix}{URL: }
\providecommand{\Pubmedprefix}{pmid:}
\providecommand{\doi}[1]{\href{http://dx.doi.org/#1}{\path{#1}}}
\providecommand{\Pubmed}[1]{\href{pmid:#1}{\path{#1}}}
\providecommand{\bibinfo}[2]{#2}
\ifx\xfnm\relax \def\xfnm[#1]{\unskip,\space#1}\fi
%Type = Article
\bibitem[{Cousins et~al.(2010)Cousins, Hymes, and Tucker}]{Cousins:2009kz}
\bibinfo{author}{R.~D. Cousins}, \bibinfo{author}{K.~E. Hymes},
  \bibinfo{author}{J.~Tucker},
\newblock \bibinfo{title}{{Frequentist evaluation of intervals estimated for a
  binomial parameter and for the ratio of Poisson means}},
\newblock \bibinfo{journal}{Nucl. Instrum. Meth. A} \bibinfo{volume}{612}
  (\bibinfo{year}{2010}) \bibinfo{pages}{388--398}.
%Type = Article
\bibitem[{Clopper and Pearson(1934)}]{10.1093/biomet/26.4.404}
\bibinfo{author}{C.~J. Clopper}, \bibinfo{author}{E.~S. Pearson},
\newblock \bibinfo{title}{{The use of confidence or fiducial limits illustrated
  in the case of the binomial}},
\newblock \bibinfo{journal}{Biometrika} \bibinfo{volume}{26}
  (\bibinfo{year}{1934}) \bibinfo{pages}{404--413}.
%Type = Article
\bibitem[{Wilson(1927)}]{doi:10.1080/01621459.1927.10502953}
\bibinfo{author}{E.~B. Wilson},
\newblock \bibinfo{title}{Probable inference, the law of succession, and
  statistical inference},
\newblock \bibinfo{journal}{Journal of the American Statistical Association}
  \bibinfo{volume}{22} (\bibinfo{year}{1927}) \bibinfo{pages}{209--212}.
%Type = Article
\bibitem[{Efron and Tibshirani(1986)}]{Efron:1986hys}
\bibinfo{author}{B.~Efron}, \bibinfo{author}{R.~Tibshirani},
\newblock \bibinfo{title}{{An introduction to the bootstrap}},
\newblock \bibinfo{journal}{Statist. Sci.} \bibinfo{volume}{57}
  (\bibinfo{year}{1986}) \bibinfo{pages}{54--75}.
%Type = Article
\bibitem[{James and Roos(1975)}]{James:1975dr}
\bibinfo{author}{F.~James}, \bibinfo{author}{M.~Roos},
\newblock \bibinfo{title}{{Minuit: A System for Function Minimization and
  Analysis of the Parameter Errors and Correlations}},
\newblock \bibinfo{journal}{Comput. Phys. Commun.} \bibinfo{volume}{10}
  (\bibinfo{year}{1975}) \bibinfo{pages}{343--367}.
%Type = Article
\bibitem[{Jeffreys(1946)}]{doi:10.1098/rspa.1946.0056}
\bibinfo{author}{H.~Jeffreys},
\newblock \bibinfo{title}{An invariant form for the prior probability in
  estimation problems},
\newblock \bibinfo{journal}{Proceedings of the Royal Society of London. Series
  A. Mathematical and Physical Sciences} \bibinfo{volume}{186}
  (\bibinfo{year}{1946}) \bibinfo{pages}{453--461}.
%Type = Misc
\bibitem[{Lam et~al.(2021)}]{siu_kwan_lam_2021_5524874}
\bibinfo{author}{S.~K. Lam}, et~al., \bibinfo{title}{numba/numba: Version
  0.54.0}, \bibinfo{year}{2021}. \DOIprefix\doi{10.5281/zenodo.5524874}.
%Type = Article
\bibitem[{Brun and Rademakers(1997)}]{Brun:1997pa}
\bibinfo{author}{R.~Brun}, \bibinfo{author}{F.~Rademakers},
\newblock \bibinfo{title}{{ROOT -- An object oriented data analysis
  framework}},
\newblock \bibinfo{journal}{Nucl. Instrum. Meth.} \bibinfo{volume}{A389}
  (\bibinfo{year}{1997}) \bibinfo{pages}{81}.
%Type = Article
\bibitem[{Harris et~al.(2020)}]{Harris:2020xlr}
\bibinfo{author}{C.~R. Harris}, et~al.,
\newblock \bibinfo{title}{{Array programming with NumPy}},
\newblock \bibinfo{journal}{Nature} \bibinfo{volume}{585}
  (\bibinfo{year}{2020}) \bibinfo{pages}{357}.
%Type = Article
\bibitem[{Virtanen et~al.(2020)}]{Virtanen:2019joe}
\bibinfo{author}{P.~Virtanen}, et~al.,
\newblock \bibinfo{title}{{SciPy 1.0: Fundamental algorithms for scientific
  computing in Python}},
\newblock \bibinfo{journal}{Nat. Methods} \bibinfo{volume}{17}
  (\bibinfo{year}{2020}) \bibinfo{pages}{261}.
%Type = Misc
\bibitem[{Dembinski et~al.(2021)}]{Dembinski:2020a}
\bibinfo{author}{H.~Dembinski}, et~al., \bibinfo{title}{scikit-hep/iminuit:
  v2.8.2}, \bibinfo{year}{2021}. \DOIprefix\doi{10.5281/zenodo.5203300}.
%Type = Article
\bibitem[{Hunter(2007)}]{Hunter:2007ouj}
\bibinfo{author}{J.~D. Hunter},
\newblock \bibinfo{title}{{Matplotlib: A 2D graphics environment}},
\newblock \bibinfo{journal}{Comput. Sci. Eng.} \bibinfo{volume}{9}
  (\bibinfo{year}{2007}) \bibinfo{pages}{90}.
%Type = Article
\bibitem[{Meurer et~al.(2017)}]{Meurer:2017yhf}
\bibinfo{author}{A.~Meurer}, et~al.,
\newblock \bibinfo{title}{{SymPy: Symbolic computing in Python}},
\newblock \bibinfo{journal}{PeerJ Comput. Sci.} \bibinfo{volume}{3}
  (\bibinfo{year}{2017}) \bibinfo{pages}{e103}.

\end{thebibliography}
\end{document}